\documentclass[12pt]{article}

\emergencystretch=\maxdimen
\usepackage{amsmath}
\usepackage{amssymb}
\usepackage{enumitem}
\usepackage{tikz}
	\usetikzlibrary{arrows.meta}
	\usetikzlibrary{decorations.markings,decorations.pathreplacing,angles,quotes,patterns,}
\usepackage{xcolor}
\usepackage{array}		
	\newcolumntype{R}[1]{>{\raggedleft\let\newline\\\arraybackslash\hspace{0pt}}m{#1}}
	\newcolumntype{L}[1]{>{\raggedright\let\newline\\\arraybackslash\hspace{0pt}}m{#1}}
	\newcolumntype{C}[1]{>{\centering\let\newline\\\arraybackslash\hspace{0pt}}m{#1}}
\usepackage{graphicx}
\usepackage{subcaption}
\def\({\left(}
\def\){\right)}
\def\[{\left[}                                
\def\]{\right]}
\newcommand{\etatype}[1]{\widetilde{#1}(\eta)}
\newcommand{\etatypesq}[1]{\widetilde{#1}^{2}\hspace{-0.014cm}(\eta)}
\newcommand{\etatypeZ}[1]{\widetilde{#1}(\eta_{0})}
\newcommand{\etatypesqZ}[1]{\widetilde{#1}^{2}\hspace{-0.014cm}(\eta_{0})}
\newcommand{\etatypecube}[1]{\widetilde{#1}^{3}\hspace{-0.014cm}(\eta)}
\newcommand{\etatypecubeZ}[1]{\widetilde{#1}^{3}\hspace{-0.014cm}(\eta_{0})}

\newcommand{\etatypep}[1]{\widetilde{#1}'\hspace{-0.014cm}(\eta)}
\newcommand{\etatypepp}[1]{\widetilde{#1}''\hspace{-0.014cm}(\eta)}
\newcommand{\ttype}[1]{{#1}(t)}
\newcommand{\ttypesq}[1]{{#1}^{2}(t)}
\newcommand{\ttypefour}[1]{{#1}^{4}(t)}
\newcommand{\ttypedot}[1]{\dot{#1}(t)}
\newcommand{\ttypedotsq}[1]{\dot{#1}^{2}(t)}
\newcommand{\ttypeddot}[1]{\ddot{#1}(t)}
\newcommand{\g}{g}
\newcommand{\flds}{\Phi}
\newcommand{\T}{T}
\newcommand{\TT}{\mathcal{T}}
\newcommand{\EIN}{\text{Ein}}
\newcommand{\real}{\mathbb{R}}
\newcommand{\tensor}{\otimes}
\newcommand{\M}{\mathcal{M}}
\newcommand{\Rint}{\mathcal{I}}
\newcommand{\ProcaA}{\mathbb{A}}
\newcommand{\ProcaF}{\mathbb{F}}
\newcommand{\ProcaC}{\Gamma}
\newcommand{\NewCoupling}{\Gamma_{1}}

\newcommand{\Kretschmann}{\mathcal{K}}
\newcommand{\Thetat}{\wt{\Theta}}

\newcommand{\MaxA}{A}
\newcommand{\MaxF}{F}

\newcommand{\Upot}{\mathcal{U}}
\newcommand{\Hub}[1]{\mathcal{H}^{#1}(t)}
\newcommand{\MagHelM}{H}
\newcommand{\MagHelP}{\mathbb{H}}
\newcommand{\tempT}{\textsf{T}}
\newcommand{\obsnu}{\nu_{\text{o}}[V_{\text{o}}]}
\newcommand{\emitnu}{\nu_{\text{e}}[V_{\text{e}}]}

\newcommand{\ScalarField}{\Psi} 

\newcommand{\indexProcaA}[1]{\ProcaA^{(#1)}}

\newcommand{\indexMaxA}[1]{\MaxA^{(#1)}}
\newcommand{\indexMaxF}[1]{\MaxF^{(#1)}}
\newcommand{\hindexMaxF}[1]{\wh{\MaxF}^{(#1)}}
\newcommand{\indexScalarField}[1]{\ScalarField^{(#1)}}

\newcommand{\w}{\wedge}
\newcommand{\SEM}{stress-energy-momentum tensor }
\newcommand{\CS}{\mathcal{R}}
\newcommand{\Lie}{\mathcal{L}}
\newcommand{\wh}[1]{\widehat{#1}}
\newcommand{\wt}[1]{\widetilde{#1}}
\newcommand{\Vol}{\text{Vol}}

\newcommand\SCALEFACVARPARAMS[2]{
	\begin{figure}[!ht]
		\centering
		\includegraphics[width=#1\textwidth]{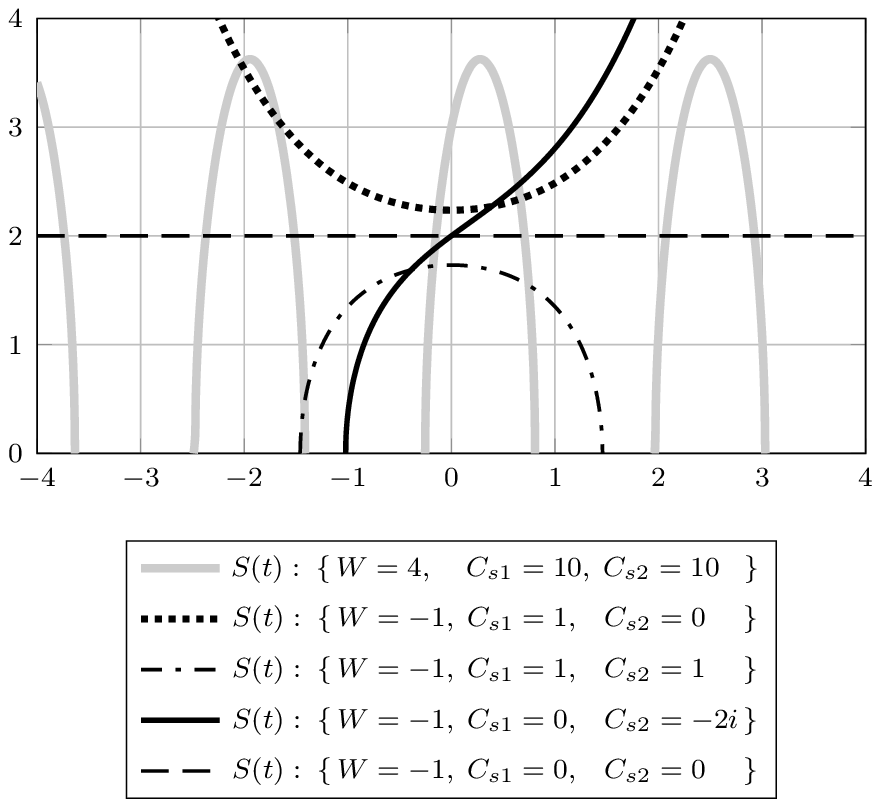}
		\caption{ #2 }
		\label{fig:scalefacvarparams}
	\end{figure}
	}
\newcommand\OBSERVDATAPLOTS[2]{
	\begin{figure}[!ht]
		\centering
		\includegraphics[width=#1\textwidth]{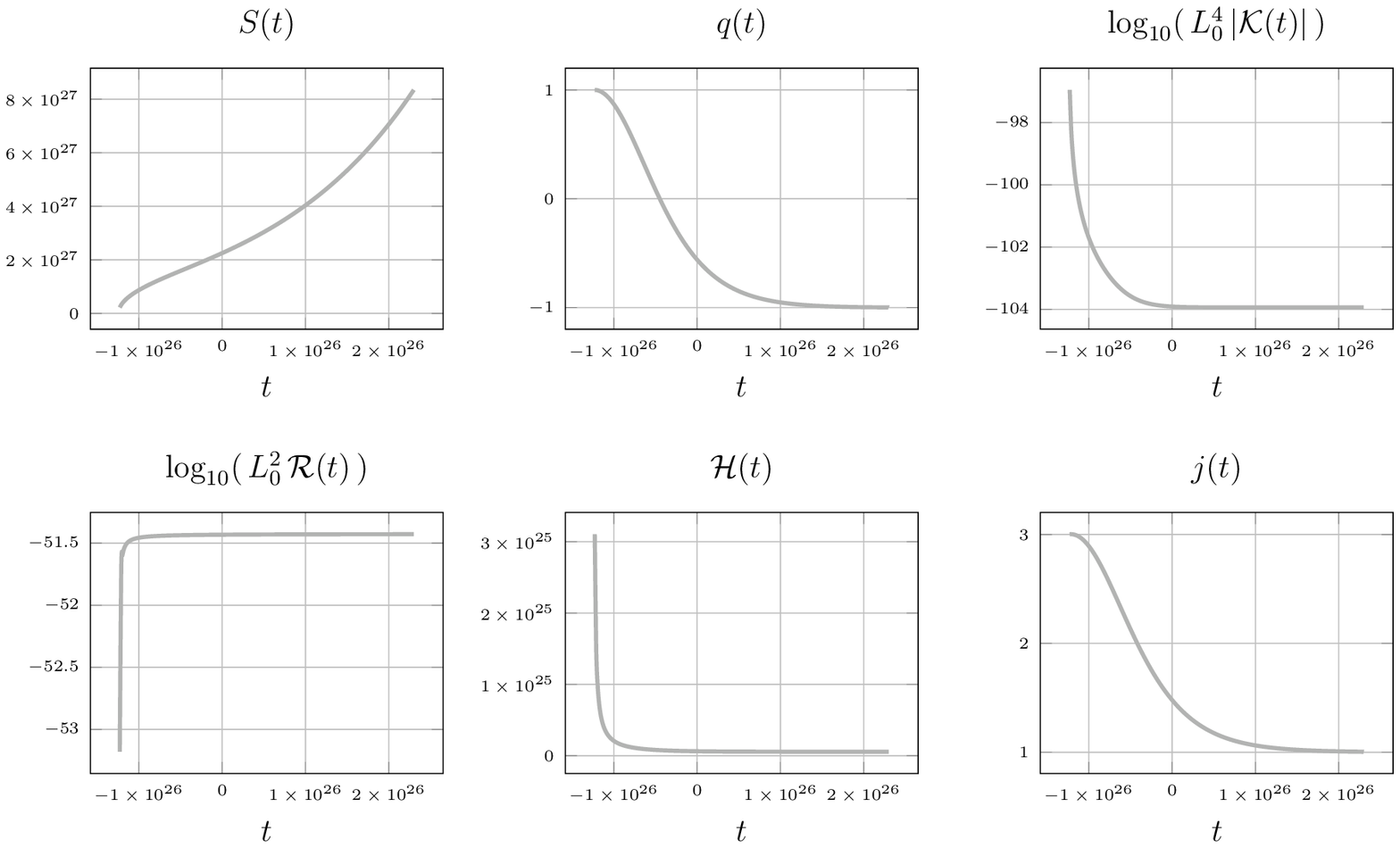}
		\caption{ #2 }
		\label{fig:observdataplots}
	\end{figure}
	}
\newcommand\BESSELFOURIERMODEDOMAINS[2]{
	\begin{figure}[!ht]
		\centering
		\vspace{0.3cm}
		\includegraphics[width=#1\textwidth]{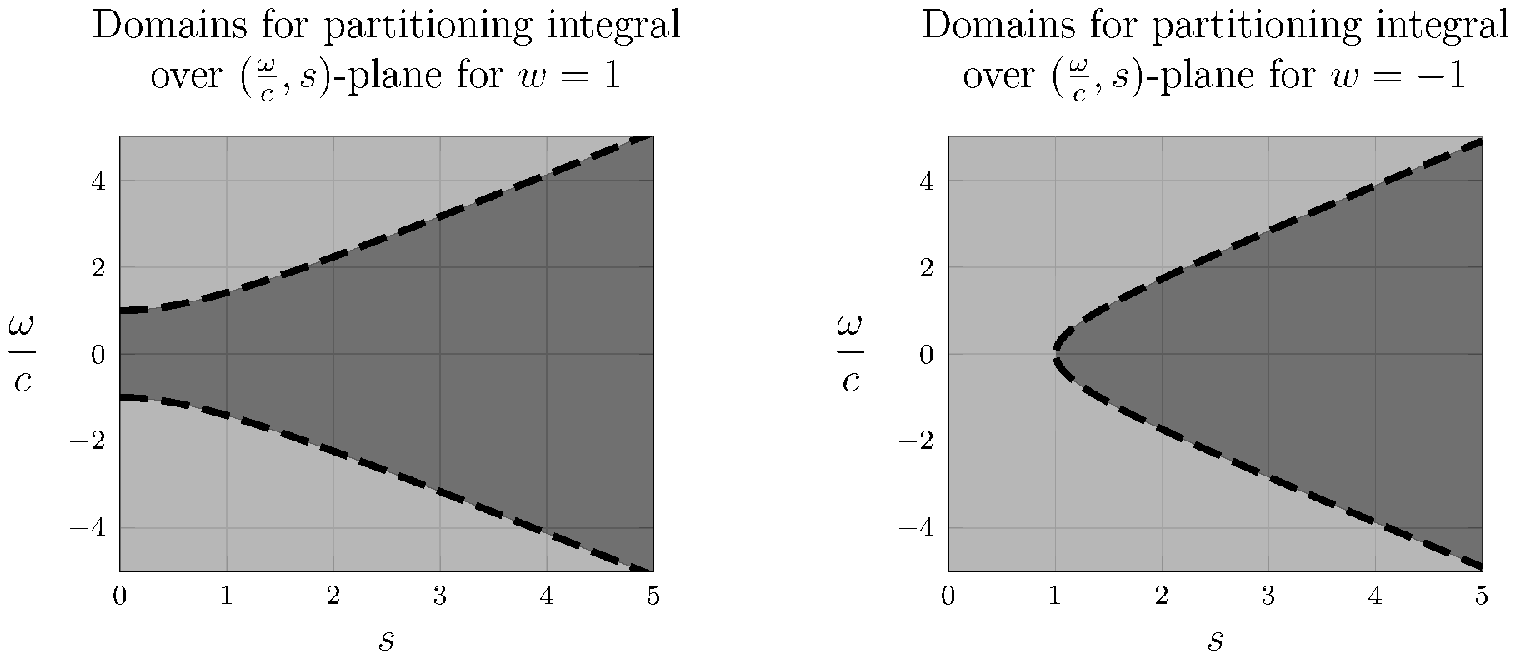}
		\caption{ #2 }
		\label{fig:besselfouriermodedomains}
		\vspace{0.3cm}
	\end{figure}
	}
\newcommand\PROCAMASSPLOTS[2]{
	\begin{figure}[!ht]
		\centering
		\includegraphics[width=#1\textwidth]{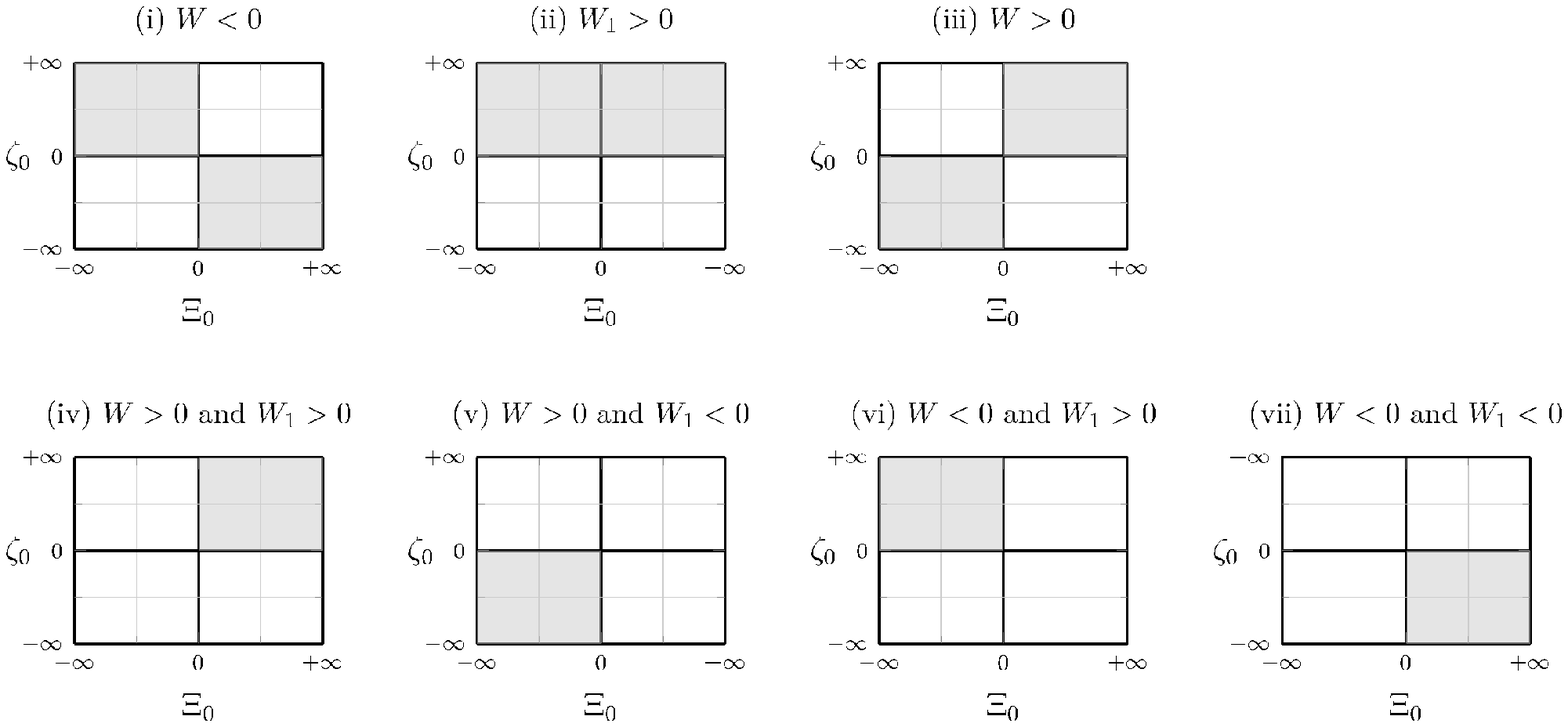}
		\caption{ #2 }
		\label{fig:procamassplots}
	\end{figure}
	}
\newcommand\ENERGYCONDS[1]{
	\begin{figure}[!p]
		\centering
		\begin{tabular}[b]{rl}
			\begin{minipage}{0.45\textwidth}
				\begin{eqnarray*}
					 \parbox{2cm}{\centering Dominant Energy Conditions} & %
												\left\{	\begin{array}{l}
													\ttype{\wh{\Omega}} \,\geq\, 0 \\[0.3cm]
													\ttype{\wh{\Omega}} - \left|\ttype{\wh{P}_{1}}\right|\,\geq\, 0 \\[0.3cm]
													\ttype{\wh{\Omega}} - \left|\ttype{\wh{P}_{2}}\right|\,\geq\, 0 \\[0.3cm]
													\ttype{\wh{\Omega}} - \left|\ttype{\wh{P}_{3}}\right|\,\geq\, 0 
												\end{array}\right.
				\end{eqnarray*} \quad \\
			\end{minipage} &	\begin{minipage}{0.5\textwidth}
									\flushright
									\includegraphics[width=\textwidth]{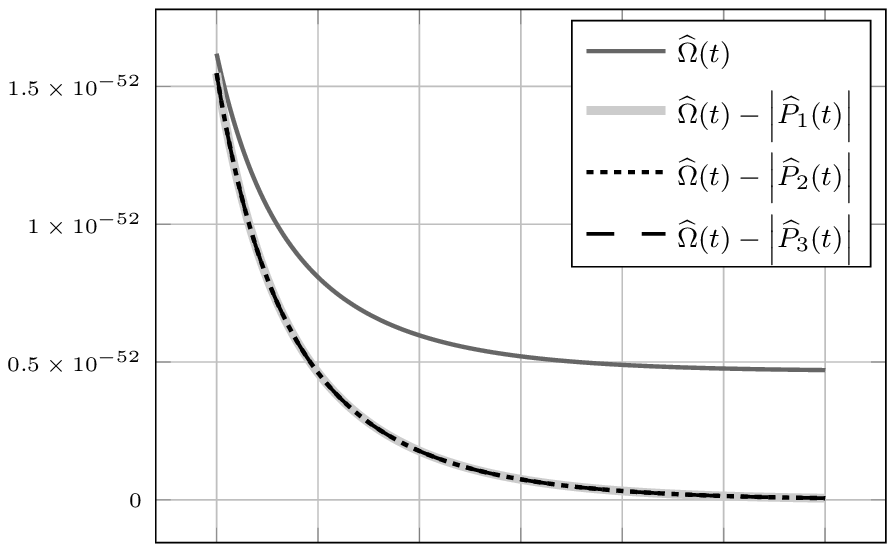}
								\end{minipage}
		\end{tabular} \quad \\[-0.2cm]    
		\begin{tabular}[b]{rl}
			\begin{minipage}{0.45\textwidth}
				\begin{eqnarray*}
					 \parbox{2cm}{\centering Strong Energy Conditions} & %
												\left\{	\begin{array}{l}
													\ttype{\wh{\Omega}} + \sum_{i=1}^{3}\ttype{\wh{P}_{i}}\,\geq\, 0 \\[0.3cm]
													\ttype{\wh{\Omega}} + \ttype{\wh{P}_{1}} \,\geq\, 0  \\[0.3cm]
													\ttype{\wh{\Omega}} + \ttype{\wh{P}_{2}} \,\geq\, 0 \\[0.3cm]
													\ttype{\wh{\Omega}} + \ttype{\wh{P}_{3}} \,\geq\, 0
												\end{array}\right.
				\end{eqnarray*} \quad \\
			\end{minipage} &	\begin{minipage}{0.5\textwidth}
									\flushright
									\includegraphics[width=\textwidth]{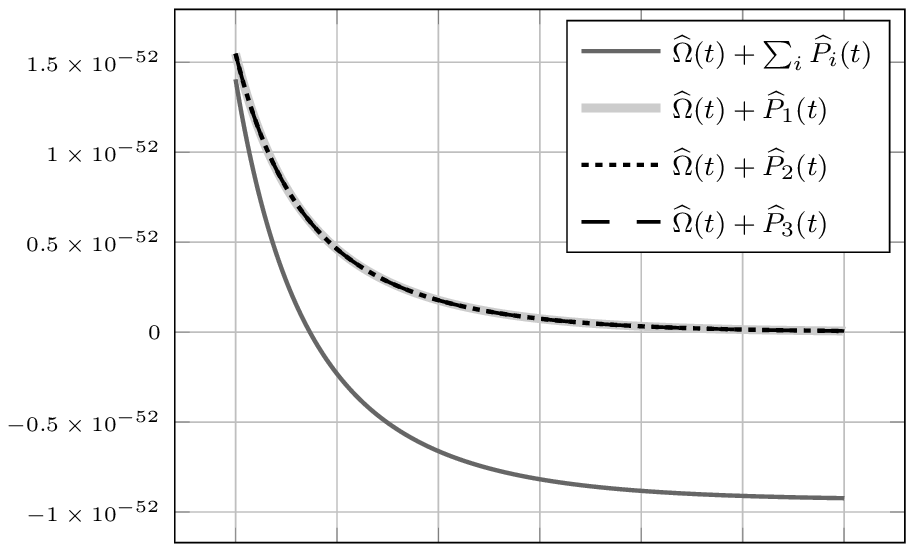}
								\end{minipage}
		\end{tabular} \quad \\[-0.2cm]  
		\begin{tabular}[b]{rl}
			\begin{minipage}{0.45\textwidth}
				\begin{eqnarray*}
					 \parbox{2cm}{\centering Weak Energy Conditions} & %
												\left\{	\begin{array}{l}
													\ttype{\wh{\Omega}} \,\geq\, 0 \\[0.3cm]
													\ttype{\wh{\Omega}} + \ttype{\wh{P}_{1}}\,>\, 0 \\[0.3cm]
													\ttype{\wh{\Omega}} + \ttype{\wh{P}_{2}} \,>\, 0 \\[0.3cm]
													\ttype{\wh{\Omega}} + \ttype{\wh{P}_{3}} \,>\, 0
												\end{array}\right.
				\end{eqnarray*} \quad \\
			\end{minipage} &	\begin{minipage}{0.5\textwidth}
									\flushright
									\includegraphics[width=\textwidth]{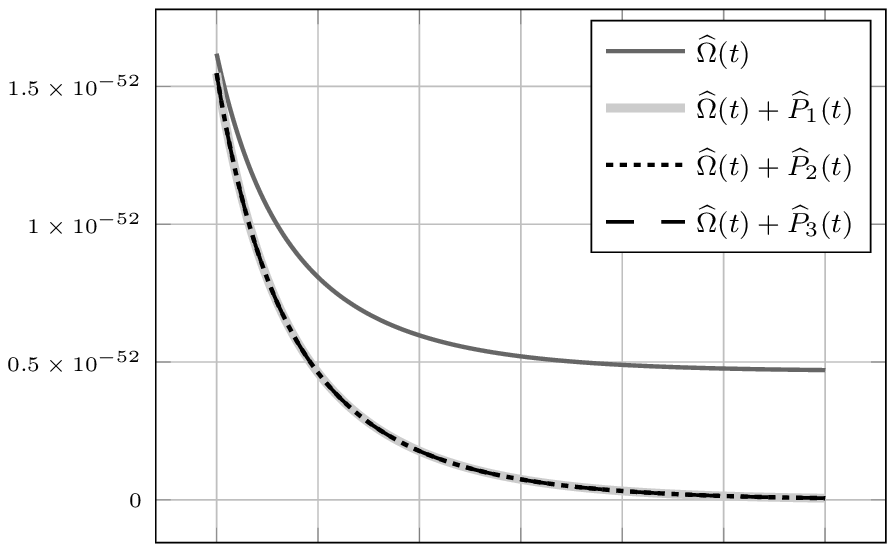}
								\end{minipage}
		\end{tabular}  \quad \\[-0.2cm]  
		\begin{tabular}[b]{rl}
			\begin{minipage}{0.45\textwidth}
				\begin{eqnarray*}
					 \parbox{2cm}{\centering Null Energy Conditions} & %
												\left\{	\begin{array}{l}
													\ttype{\wh{\Omega}} + \ttype{\wh{P}_{1}}\,\geq\, 0 \\[0.3cm]
														\ttype{\wh{\Omega}} + \ttype{\wh{P}_{2}}\,\geq\, 0 \\[0.3cm]
														\ttype{\wh{\Omega}} + \ttype{\wh{P}_{3}} \,\geq\, 0 \\[0.3cm]
												\end{array}\right.
				\end{eqnarray*} \quad \\
			\end{minipage} &	\begin{minipage}{0.5\textwidth}
									\includegraphics[width=\textwidth]{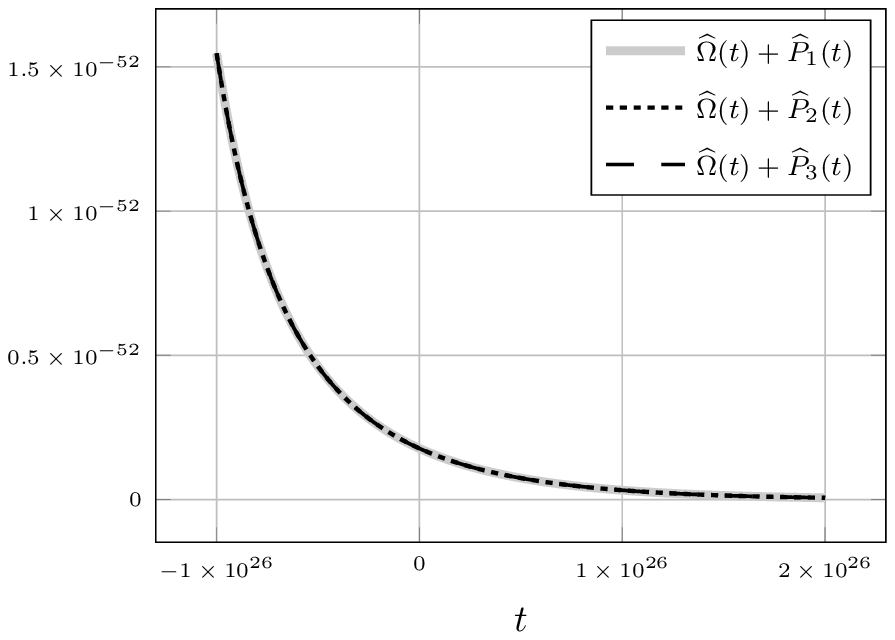}
								\end{minipage}
		\end{tabular}
		\caption{#1}
		\label{fig:energyconds}
	\end{figure}
	}

\title{A new paradigm for the dynamics of the early Universe}
\author{\begin{tabular}{R{7.9cm}C{0.2cm}L{7cm}} 
			R. W. Tucker  &	&  T. J. Walton \\[-0.4cm]
			\parbox[t]{7.9cm}{\small \linespread{1.2} \raggedleft Department of Physics, University of Lancaster and Cockcroft Institute, Daresbury Laboratory, Warrington, UK} &  & \parbox[t]{7cm}{\small \linespread{1.2} \raggedright Department of Mathematics, University of Bolton, Deane Campus, Bolton, UK} \\[1cm]
			M. Array\'{a}s	&	& J. L. Trueba \\[-0.4cm]
			\multicolumn{3}{c}{{\small  \'{A}rea de Electromagnetismo, Universidad Rey Juan Carlos, Madrid, Spain}}
		\end{tabular}
		}

\date{}
\begin{document}
\maketitle			
\begin{abstract}
	This paper invokes a new mechanism for reducing  a coupled system of fields (including Einstein's equations without a cosmological constant) to equations that possess solutions exhibiting characteristics of immediate relevance to current observational astronomy.
	
	Our approach is formulated as a classical Einstein-vector-scalar-Maxwell-fluid  field theory on  a spacetime with three-sphere spatial sections.  Analytic  cosmological solutions are found  using local charts familiar from standard LFRW cosmological models. These solutions can be used to describe different types of evolution for the metric scale factor, the Hubble,  jerk and de-acceleration functions, the scalar spacetime curvature and the Kretschmann invariant constructed from the Riemann-Christoffel spacetime curvature tensor. The cosmological sector of the theory accommodates a particular single big-bang scenario followed by an eternal exponential acceleration of the scale factor. Such a solution does {\it not} require an externally prescribed fluid equation of state and leads to a number of new predictions including  a current value of the  ``jerk'' parameter,  ``Hopfian-like''   source-free Maxwell field configurations with magnetic helicity and distributional ``bi-polar'' solutions exhibiting a new charge conjugation symmetry.
	
	An approximate scheme for field perturbations about this particular cosmology is explored and its consequences for a thermalisation process and a thermal history are derived, leading to a prediction of the time interval between the big-bang and the decoupling era. Finally it is shown that field couplings exist where both vector and scalar localised linearised perturbations exhibit dispersive wave-packet behaviours.  The scalar perturbation may also give rise to Yukawa solutions associated with a massive Klein-Gordon particle. It is argued that the vector and scalar fields may offer candidates for ``dark-energy'' and ``dark-matter'' respectively.\\[0.3cm]
	
\end{abstract}
\vspace{2pc}

\section*{Introduction}
The advent of modern satellite technology in observational astronomy has ushered in a new era  for astrophysics and research into the fundamental role of gravitation on a large scale. In particular the venerable subject of relativistic cosmology has received a new impetus with the realisation that  a number of standard cosmological models  may need revision \cite{ijjas_pop,ijjas_inflation}. In this paper we discuss an alternative paradigm that, while retaining  many  of the most established features of the current standard model, circumvents some of its weaknesses.  We invoke a new mechanism for reducing Einstein's field equations (without a cosmological constant  term)  coupled to fluid matter,  vector and  scalar fields,  to a dynamical system  that possesses a class of  {\it simple analytic}  solutions  for the metric scale factor exhibiting characteristics of immediate relevance to current observational astronomy.\\

Within the framework of the standard cosmological paradigm there has long been interest in models containing additional vector and/or scalar fields. One that has had considerable impact due to its applications to astrophysics and the problem of hidden matter was pioneered by Milgrom \cite{milgrom} and developed extensively by Bekenstein \cite{bekenstein}. Bekenstein's model offers a relativistic post-MOND theory designed to account for a broad range of astrophysical phenomena, including galactic rotation rates, without explicitly  introducing {\it cosmological dark-matter fields}. It is constructed in terms of a constrained dynamic vector field ($\Pi$), a dynamic tensor field ($\g$), a dynamic scalar field ($\phi$), a non-dynamic field ($\sigma$), a Lagrange multiplier field ($\lambda$), a phenomenological real-valued function ($F$) and four coupling parameters $K,k,l,G$. Matter is introduced following the standard paradigm in terms of ideal multi-fluids, each with their individual equations of state. It is further assumed that each fluid 4-velocity is aligned with the time-like vector field $\Pi$ and that the `physical metric' in the fluid stress-energy-momentum tensors is non-trivially related to the tensor field $\g$.  Application of the theory to astrophysical phenomena demands wide ranging  assumptions and approximations.  Although it claims agreement with extra-galactic phenomena, including the lensing of electromagnetic radiation by galaxies and galaxy clusters, concordance with the solar system and binary pulsar tests, it leaves open the question of the need for {\it cosmological dark matter}.\\

Our model has more modest aims. It constructs a viable cosmology in terms of four dynamical fields $\ProcaA,\MaxF,\ScalarField$, the physical metric tensor $\g$ and fewer coupling constants. No phenomenological function is involved and the presence of the Maxwell field $\MaxF$ is necessary for developing the electromagnetic sector and essential in identifying the fluctuation solutions $\indexProcaA{1}, \indexScalarField{1}$ as {\it dark} elements. The Bekenstein theory follows the standard paradigm of solving the homogeneous, isotropic Einstein equation, involving a multi-fluid with prescribed individual equations of state, for the scale factor. By contrast we propose an alternative paradigm by exploiting an anisotropic ans\"{a}tz for the vector field $\ProcaA$ and a constant value for the scalar field $\ScalarField$ leading to a dynamic equation for the scale factor. It remains for future work to ascertain  all values of the constants in our theory and to verify whether such a minimalist field system is consistent with the available astrophysical data beyond that considered in this paper.\\

In section 1 we briefly outline some aspects of the standard approach and draw attention to those weaknesses that have motivated our approach, described in section 2.  In section 3 we introduce  our general model in terms of a system of field equations for a scalar and vector field interacting with Maxwellian electromagnetism, an anisotropic material fluid and Einsteinian gravitation  on spacetime. To develop a particular dynamical cosmological model from this general model we exploit the  Maurer-Cartan  group structure of the 3-sphere ($S^3$) to construct a preferred spacetime frame and spacetime metric  that shares the homogeneity and isotropy characteristics possessed by those  Lema\^{i}tre-Friedmann-Robertson-Walker (LFRW)  metrics with a closed spatial topology.\\

In section 4 we explicitly construct a class of general analytic solutions for this dynamical system using local charts on spacetime familiar from the  LFRW  cosmological models. In particular charts,  solutions for the LFRW scale factor  involve only bounded or hyperbolic trigonometric functions, square roots, a single real parameter  and a pair of arbitrary  integration constants. We illustrate  the different types of  Universe evolution that  may arise with particular reference to the cosmic time-dependence of the metric scale factor, the Hubble and de-acceleration parameters, the scalar spacetime curvature, the Kretschmann scalar invariant (constructed from the Riemann-Christoffel spacetime curvature tensor and the spacetime metric) once these constants are fixed.  We then exhibit the solutions to the Einstein equations  for the fluid variables  in terms of the scale factor and solutions for the remaining fields in the system. A notable feature  of this paradigm is that no additional auxiliary conditions (such as fluid equations of state or Vlasov models) are imposed. By contrast, having fixed all couplings and constants of integration, all cosmological variables  have a unique evolution and the fluid variables satisfy  algebraic constraints. We interpret such constraints as  dynamically induced fluid anisotropic equations of state.  Furthermore, because the three-dimensional spatial sections of spacetime with constant time are closed in this model, we show that the vector fields each give rise to configurations with finite total  {\it instantaneous magnetic helicity}.  For the Maxwell field this offers a possible  interpretation of primordial magnetic fields \cite{thorne_primordial} with closed knotted field lines in an expanding Universe.\\

In section 5 we use a recent value of the Hubble parameter to single out a particular solution compatible with the current estimate of the {\it negative} de-acceleration parameter and the existence of an initial singular state  for the Universe.  The model exhibits  a non-uniform exponential expansion for the LFRW scale factor into the future, predicts an age for the Universe and a current value for the unobserved {``jerk''} parameter \cite{visser_jerk}.\\

In section 6 we explore the nature of other solutions by  field perturbations about the cosmological solution and show how they may give rise to localised electrically charged domains in space as a prerequisite for scattering processes leading to a thermalisation mechanism. The data  used in the cosmological model together with the Planck spectrum is then employed in section 7  to discuss   a predicted thermal history  leading to an estimate of the time between the big-bang and the decoupling era.  Since the scalar and vector field perturbations  may provide observational signatures,  in section 8 we analyse their first-order partial differential equations (PDEs) for spatially localised solutions in domains  where the spacetime is approximately flat. We conclude that the scalar perturbation may  give rise to the existence of a massive classical Klein-Gordon particle and both perturbations admit localised dispersive wave-packet solutions.\\
 
The concluding section summarises the essential features discussed in the paper.

\section{Standard Cosmological Models}
The standard cosmological paradigm is modelled on a class of spacetimes that are spatially homogeneous and isotropic \cite{plebanski_GRbook,weinberg_cosmology}. This class is labelled by an index ($k=0,\pm 1$) describing spatial global topology with associated symmetry properties and leads to a similarly labelled class of spacetime metrics $\{\g_{(k)}\}$. Each metric is supplemented by a choice of a symmetric second-rank tensor $\T_{(k)}$ used in Einstein's field equations:
\begin{eqnarray}\label{EFE}
		\kappa\,\EIN[\g_{(k)}] + \Lambda_{0}\,\g_{(k)} \,=\, \T_{(k)} 
\end{eqnarray}
where $\Lambda_{0}\in\real$ denotes the ubiquitous cosmological constant and $\kappa = c^{4}/8\pi G$ in terms of the speed of light in vacuo $c$ and the Newtonian gravitational constant $G$. The Einstein tensor $\EIN[\g_{(k)}]$  is defined in terms of the Levi-Civita connection $\nabla$ calculated from  $\g_{(k)}$ where $\EIN[\g_{(k)}]=\text{Ric} - \frac{1}{2}\CS\g_{(k)}$ in terms of the  associated Ricci tensor $\text{Ric}$ and Ricci curvature scalar $\CS$ associated with $\g_{(k)}$. Each $ \T_{(k)}$ takes the form:
\begin{eqnarray}\label{StandardT}
	\T_{(k)} \,=\, \sum_{r=1}^{N} \,\TT^{(r)}_{(k)}[\g_{(k)},\rho_{(r)}, p_{(r)}(\rho_{(r)})]
\end{eqnarray}
where $\TT^{(r)}_{(k)}$ often takes the {\it perfect} fluid form:
\begin{eqnarray*}
	\TT^{(r)}_{(k)} \,=\, \(\,\rho_{(r)}c^{2} + p_{(r)}(\rho_{(r)})\,\)V \tensor V + p_{(r)}(\rho_{(r)})\,\g_{(k)}
\end{eqnarray*}
for some preferred vector field $V$ on spacetime and partial pressure $p_{(r)}(\rho_{(r)})$ specified in terms of some partial mass-energy density $\rho_{(r)}c^{2}$. Since  $\nabla\cdot\EIN[\g_{(k)}]=\nabla\cdot \g_{(k)}=0$  identically  in Einstein's theory of gravitation and the  $N$ fluids are non-interacting (i.e.  each $\TT^{(r)}_{(k)}$ only depends upon $\rho_{(r)}$), the necessary (Bianchi) consistency condition      $\nabla\cdot \T_{(k)}=0$ for all solutions of (\ref{EFE})  to satisfy can be achieved by imposing the  individual conditions: 
\begin{eqnarray}\label{StandardPartialDiv}
	\nabla\cdot \TT^{(r)} _{(k)}\,=\, 0 \qquad (\text{for}\quad r=1,2,\ldots,N).
\end{eqnarray} 
Many consequences of the standard cosmological paradigm follow from such impositions which are by no means fundamental since  the  only necessary condition  here is 
\begin{eqnarray*}
	\nabla\cdot \T_{(k)} \,=\, 0 .
\end{eqnarray*} 
From the assumed homogeneity and spatial isotropy for the model, $\g_{(k)}$ can be parametrized by a single real positive scalar function $S_{(k)}:\real\rightarrow\real$ for each $k$, which is a function of some evolution variable in a local spacetime chart. The vector field $V$ is then chosen to be the common unit time-like flow for the composite fluid described by the tensor $\T_{(k)}$. Different equations of state  for each  $r$ are assumed to dominate different epochs in the evolution of the Universe from some initial state to its current state. This evolution is governed by (\ref{EFE}) which reduces to a system of independent ordinary differential equations (ODEs) involving $\{S_{(k)},\rho_{(r)}\}$, given values for the constants  $\{\Lambda_{0},N\}$, initial values for $S_{(k)}$ and its first derivative, and the $r$ equations of state $p_{(r)}(\rho_{(r)})$. Equation (\ref{StandardPartialDiv}) can, in principal, be integrated to yield $\rho_{(r)}$ in terms of $S_{(k)}$ at the expense of introducing an additional  constant of integration for each $r$. \\

This program has been pursued extensively in the past in attempts to pin down the appropriate class index $k$ based upon astrophysical data for $\rho_{(r)}$, current values for $S_{(k)}$ and its rate of change. Furthermore, the discovery of (almost) isotropic black-body thermal electromagnetic radiation has led some authors \cite{guth_inflation} to invoke additional fields (inflatons) to modify the dynamics described by (\ref{EFE}) and (\ref{StandardPartialDiv}). Further observations of recent supernova luminosities have raised further concerns that the material content of the Universe, based upon current values of $\rho_{(r)}$, cannot account for the observational data for the current  de-acceleration parameter calculated from  the variation of  $S_{(k)}$ with epoch. This has led to a number of alternative choices for the equations of state $p_{(r)}(\rho_{(r)})$ and the retention of the cosmological constant $\Lambda_{0}$, sometimes interpreted as a ``Casimir energy'' of space. 

\section{An Overview of the New Paradigm}
In view of these perceived shortcomings we explore an alternative paradigm that {\it maintains the basic premise of homogeneity and isotropy for the spacetime metric in the model} but dispenses with the imposition of any supplementary fluid equations of state. Instead, we invoke a general model involving field equations for a Maxwell field and a vector and scalar field with specific couplings to each other and gravity, in addition to Einstein's equations involving a material anisotropic fluid. We declare at the outset that the fully coupled system admits a class of $k=+1$ LFRW-type cosmologies given a particular ans\"{a}tz for the solutions to these field equations. \\

We assume throughout that the Universe is described by a triple $(\M,\g,\flds)$ where $\M$ is a spacetime endowed with a metric $\g$ and $\flds$ is a collection of classical fields. It is implicit in our paradigm that:
\begin{quote}
	Spacetime $\M=\Rint\times S^{3}$ where $\Rint$ denotes an interval of the real line coordinated by $t$ and $S^{3}$ is topologically a 3-sphere. The manifold $\M$ is endowed with a $k=+1$ isotropic LFRW covariant Lorentzian metric tensor $\g$ in terms of a scale factor $S(t)$ with $\flds$ comprised of a real scalar field $\ScalarField$, a real 1-form $\ProcaA$ and a real Maxwell 2-form $\MaxF$.
\end{quote}
Our particular cosmological model is determined by a system of field equations of the form:
\begin{alignat*}{2}
	(\text{I}) &	\quad &\mathcal{F}[S,\dot{S},\ddot{S},\ScalarField,\ProcaA,\MaxF] &\,=\, 0 \\[0.2cm]
	(\text{II})&	\quad &\kappa\EIN[S,\dot{S},\ddot{S}] &\,=\, T[S,\ScalarField,\ProcaA,\MaxF,\rho,p_{1},p_{2},p_{3}] \,=\, \mathcal{T}[S,\ScalarField,\ProcaA,\MaxF] + \mathbb{T}[S,\rho,p_{1},p_{2},p_{3}].
\end{alignat*}
Equations (\textrm{I}) and (\textrm{II}) denote a coupled system of second-order partial differential equations involving the fields and the scale factor $S$ and equation (\textrm{II}) denotes Einstein's field equations in terms of the Einstein tensor $\EIN$ and total \SEM $T$. This is decomposed into field and matter components: $\mathcal{T}$ and $\mathbb{T}$ respectively, where $\rho c^{2}$ denotes the cosmological eigen-mass-energy density and $p_{1},p_{2},p_{3}$ the eigen-pressures of the {\it single} fluid \SEM of the matter in the Universe. The paradigm is to solve (\textrm{I}) and (\textrm{II}) as follows:
\begin{enumerate}
	\item	Construct a suitable ans\"{a}tz for the fields $\ScalarField,\ProcaA,\MaxF$; such that: 
	\item	the partial differential field equations (\textrm{I}) are satisfied with $S(t)$ a solution to a second order ordinary differential equation, thereby determining the metric;
	\item	the Einstein field equations (\textrm{II}) are then solved {\it algebraically} for the eigen-mass-energy density and eigen-pressures $\{\rho c^{2},p_{1},p_{2},p_{3}\}$. 
\end{enumerate}
An important aspect of this paradigm is that the algebraic relations for $\rho$ and $p_{1},p_{2},p_{3}$, considered as an {\it induced cosmological anisotropic mechanical equation of state}, are not postulated {\it a priori}. A consequence of these assumptions is that $\EIN[S,\dot{S},\ddot{S}]$ contains no term of the form $\Lambda_{0}\g$ with $\Lambda_{0}\in\real$. With a {\it bona fide} solution to Einstein's field equations, the {\it total} \SEM $T$ has no term of the form $\Lambda_{0}\g$, implying our model is devoid of a cosmological constant. Such terms do arise in the {\it field} \SEM $\mathcal{T}$ but are cancelled by terms in the matter \SEM $\mathbb{T}$, following from the solutions for the eigenvalues $\{\rho c^{2},p_{1},p_{2},p_{3}\}$.

\section{The General Vector-Scalar Model}
\subsection{The Master Field Equations}
We establish our general field system on the spacetime  manifold $\M=\Rint\times S^{3}$ where $\Rint$ is an open subset of the real line and $S^{3}$ is topologically a 3-sphere. We endow $\M$ with a covariant Lorentzian metric tensor $\g$, a real scalar field $\ScalarField$, a real 1-form $\ProcaA$ and a real 2-form $\MaxF$ satisfying the master field  equations:
\begin{eqnarray}
	\label{psi_eqn}		d\star d\ScalarField - \frac{1}{2}\kappa_{0}\,\Upot'(\ScalarField) \ProcaA \w \star \ProcaA + \frac{1}{2}\NewCoupling\,\Upot'(\ScalarField) \MaxF \w \MaxF \,=\, 0 \\[0.3cm]
	\label{max_eqn} 	d\MaxF \,=\, 0, \qquad d\star\MaxF + \NewCoupling\,\Upot'(\ScalarField)\, d\ScalarField \w \MaxF \,=\, 0 \\[0.5cm]
	\label{proca_eqn}	d\star d\ProcaA + \kappa_{0}\,\Upot(\ScalarField)\star \ProcaA \,=\, 0
\end{eqnarray}
where $\kappa_{0},\NewCoupling$ are  non-zero real coupling constants, with $\kappa_{0}$ having physical dimensions and $\NewCoupling$ dimensionless. The map  $\Upot:\real\rightarrow\real$ is a potential function and  $\star$ denotes  the linear Hodge map on forms associated with the general metric $\g$ on $\M$.
Since, for all {\it real}  fields   $ \ScalarField $  and $ \ProcaA  $,  equations (\ref{max_eqn}) are invariant under the transformation $ \MaxA \mapsto \MaxA + d\,\lambda$, where $\MaxF=d\MaxA$ and $\lambda$ is an arbitrary $0$-form,  we identify them with the  covariant Maxwell equations for the electromagnetic $U(1)$ gauge invariant 2-form $\MaxF$.  Their representation in the SI system will be described below when we discuss the electric current sources in  section 6. Equation (\ref{proca_eqn}) implies the integrability condition: 
\begin{eqnarray*}
	d(\,\Upot(\ScalarField)\star \ProcaA) &=& 0.
\end{eqnarray*}
Since  equations (\ref{psi_eqn}),  (\ref{max_eqn})  and (\ref{proca_eqn})  arise from taking field variations of the action 4-form on $\M$:
\begin{eqnarray}
\begin{split}\label{ACTION}
	\mathcal{S}[\ScalarField,\ProcaA,\MaxA,\g] \,=\, \frac{1}{2}d\ScalarField \w \star d\ScalarField  &+ \frac{1}{2}d\ProcaA \w \star d\ProcaA + \frac{1}{2}\kappa_{0}\;\Upot(\ScalarField)\,\ProcaA \w \star \ProcaA  \\[0.3cm]
		&+ \frac{1}{2}d\MaxA \w \star d\MaxA + \frac{1}{2}\NewCoupling \MaxA \w d\MaxA \w d(\,\Upot(\ScalarField)\, ) 
\end{split}
\end{eqnarray}
it  is straightforward to derive a  symmetric \SEM $\TT$ associated with the fields $\{\ScalarField,\ProcaA,\MaxA\}$ from its metric variations: 
\begin{eqnarray*}
		\TT &=&  \Upot(\ScalarField)\( d\ScalarField\tensor d\ScalarField - \frac{1}{2}\g(d\ScalarField,d\ScalarField)\,\g\) - \(i_{a}d\ProcaA\tensor i^{a}d\ProcaA + \frac{1}{2}\star(d\ProcaA\w\star d\ProcaA)\g\) \\[0.2cm]
		 	&&	+ \kappa_{0}\;\Upot(\ScalarField)\(\ProcaA \tensor \ProcaA - \frac{1}{2}\g(\ProcaA,\ProcaA)\,\g\) - \(i_{a}d\MaxA\tensor i^{a}d\MaxA + \frac{1}{2}\star(d\MaxA\w\star d\MaxA)\g\)
\end{eqnarray*}
where $i_{a}\equiv i_{X_{a}}$ denotes the interior contraction operator on forms with respect to any local $\g$-orthonormal basis of vector fields $\{X_{a}\}$ on $\M$ and $i^{a}\equiv \eta^{ab}\,i_{X_{b}}$ with $\eta^{ab}=\text{diag}(-1,1,1,1)$. Note that the term in (\ref{ACTION}) containing the coupling $\NewCoupling$ does not contribute to the metric variations. \\

In addition to $\TT$ we include the single material fluid symmetric tensor $\mathbb{T}$ parametrized in a local $g-$orthonormal eigen-cobasis $\{e^{a}\}$ on $\M$ as
\begin{eqnarray*}
	\mathbb{T} \,=\, \rho c^{2} e^{0}\tensor e^{0} + \sum_{k=1}^{3} p_{k}e^{k}\tensor e^{k}
\end{eqnarray*}
in terms of the real scalar field components $\{\rho,p_{1},p_{2},p_{3}\}$ on $\M$. Variational methods for deriving fluid material \SEM tensors can be found in \cite{eckart_varhydro,seliger_varCM,karlovini_stars}. The system is closed by adopting $T=\TT+\mathbb{T}$ as the source tensor for Einstein's field equations {\it without}  a cosmological constant:
\begin{eqnarray}\label{OurEFE}
	\kappa\,\EIN[\g] &=& T
\end{eqnarray}
Since $\kappa$ has physical dimensions of $\text{ML}/\text{T}^{2}$ in the SI system of units it is convenient to assign SI physical dimensions consistently to all tensors in the field system above. To this end, we assign to the covariant rank-two  spacetime metric tensor $\g$ the physical dimensions of $\text{L}^2$ and write
\begin{eqnarray}\label{metricortho}
	\g &=& -e^{0}\tensor e^{0} + \sum_{k=1}^{3}e^{k}\tensor e^{k}
\end{eqnarray}
in any $\g$-orthonormal cobase $\{e^{a}\}$ with elements having physical dimension of length: $[e^{a}]=\text{L}$. Then the dimensions of the curvature 2-forms $\{R_{ab}\}$ and the curvature scalar 
\begin{eqnarray*}
	\CS=\star\(R_{ab}\w \star(e^{a} \w e^{b})\)
\end{eqnarray*}
satisfy $[R_{ab}]=1$ and $[\CS]=\text{L}^{2}$. Furthermore, these imply $[\EIN]=1$ yielding dimension of the \SEM tensor $[T]= [\TT]+[\mathbb{T}] =  \text{ML}/\text{T}^{2}$   since $[\kappa]= \text{ML}/\text{T}^{2}  $. Bearing in mind that for any $p$-form $\beta$ on an $n$-dimensional manifold: 
$[\beta]=[d\beta]$ and
\begin{eqnarray*}
	[\star\beta]\,=\,[\beta]\,\text{L}^{n-2p},
\end{eqnarray*}
we may then consistently assign $[\ScalarField]=1$, $[\ProcaA]=[\MaxA]=\text{L}$,
\begin{eqnarray*}	
	\kappa_{0}\,=\,\frac{\zeta_{0}}{L_{0}^{2}}
\end{eqnarray*}
where $[L_{0}]=\text{L}$, $[\zeta_{0}]=1$, $[\Upot(\ScalarField)]=[\Upot'(\ScalarField)]=1$, $[\NewCoupling]=1$. Furthermore the material density $\rho$ has MKS dimensions $\text{M}/\text{L}^3$ and the pressures $p_{1},p_{2},p_{3}$  have MKS dimensions $\text{M}/\text{L}\text{T}^2$  as usual. If all parameters and tensors are specified  with values in the SI system of units then the value of the parameter $L_{0}$ is unity. \\

\subsection{The Cosmological Metric Ans\"{a}tz}
To construct a particular cosmological model based upon the system of equations (\ref{psi_eqn}), (\ref{max_eqn}), (\ref{proca_eqn}) and (\ref{OurEFE}), we need a suitable particular ans\"{a}tz for $\g,\ScalarField,\ProcaA,\MaxA$ {\it and} a preferred eigenbasis for $\mathbb{T}$ defining $\{\rho,p_{1},p_{2},p_{3}\}$. To this end, we note that $S^{3}$ may be identified with the group manifold $SO(3)$ with an algebra generated by the canonical $(2\times 2)$ Pauli matrices $\{\sigma^{1},\sigma^{2},\sigma^{3}\}$. A group element $U\in SO(3)$ can be coordinated by the three {\it{real}} numbers $\{\alpha_{1},\alpha_{2},\alpha_{3}\}$ such that:
\begin{eqnarray*}
	U \,=\, \exp\[i(\alpha_{1}\sigma^{1} + \alpha_{2}\sigma^{2} + \alpha_{3}\sigma^{3})\]	.
\end{eqnarray*}
The three pure imaginary differential 1-forms on $SO(3)$ defined by
\begin{eqnarray*}
	\Thetat^{k} \,=\, \frac{1}{2}\text{tr}(\sigma^{k}\,U^{-1}dU) \qquad (k=1,2,3)
\end{eqnarray*}
satisfy the Maurer-Cartan relations at each point with coordinates $(\alpha_{1},\alpha_{2},\alpha_{3})$:
\begin{eqnarray}\label{MCrel}
	d\Thetat^{k} \,=\, -2i\epsilon^{k}_{\;\;lm}\,\Thetat^{l} \w \Thetat^{m} \qquad (k,l,m=1,2,3)
\end{eqnarray}
where $\epsilon^{k}_{\;\;lm}$ is the Levi-Civita alternating symbol. The expressions for $\{\Thetat^{k}\}$ in terms of $\{\alpha_{1},\alpha_{2},\alpha_{3}\}$ are given in the appendix. These 1-forms offer a preferred cobasis of 1-forms on the group manifold $SO(3)\cong S^{3}$ and will be used below  to define a preferred class of metrics on $\M$. We show in the appendix that a stereographic mapping (\ref{StereoCoordT}) from the coordinates $(\alpha_{1},\alpha_{2},\alpha_{3})$ to a local coordinate chart  on $S^3$ with real coordinates $(\xi_{1},\xi_{2},\xi_{3})$ gives simpler expressions for the cobasis 1-forms on $S^{3}$, denoted $\{\Theta^{1},\Theta^{2},\Theta^{3}\}$. To define a preferred cobasis  of 1-forms $\{e^0,e^1,e^2,e^3\}$ on {\it spacetime}, we adopt a local spacetime chart with   dimensionless coordinates $\{\eta,\xi_{1},\xi_{2},\xi_{3}\}$ where $\eta\in\Rint$ and express the {\it real}  Lorentzian cosmological metric on $\M$ as (\ref{metricortho}) in terms of
\begin{eqnarray}\label{eta_cobas}
	e^{0} \,=\, L_{0}\,\etatype{S}\;d\eta\qquad\text{and}\qquad e^{k} \,=\, i\,L_{0}\,\etatype{S}\;\Theta^{k} \qquad (k=1,2,3)
\end{eqnarray}
constituting a real $\g$-orthonormal cobasis. In this spacetime chart the metric ans\"{a}tz  thereby takes the form:
\begin{eqnarray*}
	\g &=& L_{0}^{2}\,\etatypesq{S}\[ -d\eta \tensor d\eta + \frac{4}{(1+\xi_{1}^{2}+\xi_{2}^{2}+\xi_{3}^{2})^{2}}\( d\xi_{1}\tensor d\xi_{1} + d\xi_{2}\tensor d\xi_{2} + d\xi_{3}\tensor d\xi_{3}\)\].
\end{eqnarray*}
A more familiar LFRW form of the metric follows from the coordinate transformation:
\begin{eqnarray*}
	\left\{\xi_{1}\,=\, \frac{1}{2}r\sin(\theta)\cos(\phi),\quad \xi_{2}\,=\, \frac{1}{2}r\sin(\theta)\sin(\phi),\quad \xi_{3}\,=\, \frac{1}{2}r\cos(\theta)\right\}
\end{eqnarray*}
yielding
\begin{eqnarray}\label{gt}
	\g &=& L_{0}^{2}\,\etatypesq{S}\[ -d\eta \tensor d\eta + \frac{1}{(1+\frac{r^{2}}{4})^{2}}\( dr\tensor dr + r^{2}d\theta\tensor d\theta + r^{2}\sin^{2}(\theta)d\phi\tensor d\phi\)\].
\end{eqnarray}
In this chart, the usual ranges $0\leq r <\infty$, $0\leq \theta \leq 2\pi$ and $0\leq \phi < \pi$ for the polar coordinates cover $S^{3}$ minus one point. The Maurer-Cartan relations (\ref{MCrel}) on $S^{3}$ imply the following structure equations for the cobasis on $\M$:
\begin{eqnarray*}
	 de^{0} \,=\, 0 \qquad\text{and}\qquad de^{k} &=& \frac{1}{L_{0}\,\etatypesq{S}}\frac{d\etatype{S}}{d\eta}\,e^{0} \w e^{k} - \frac{2}{L_{0}\,\etatype{S}}\,\epsilon^{k}_{\;\;lm}\,e^{l} \w e^{m}
\end{eqnarray*}
for $k=1,2,3$, which greatly facilitate many of the computations below. \\

One readily confirms that 
\begin{eqnarray}\label{TLUV}
	X_{0} &=& \frac{1}{L_{0}\,\etatype{S}}\frac{\partial}{\partial\eta}
\end{eqnarray}
is a unit timelike vector field on $\M$ and that the vector fields $K_{j}=\etatype{S}\,X_{j}$ are spacelike Killing vectors on $\M$:
\begin{eqnarray*}
	\Lie_{K_{j}}\g &=& 0 \qquad (j=1,2,3)
\end{eqnarray*}
where $e^{a}(X_{b})=\delta^{a}_{b}$ with $a,b=0,1,2,3$. The vector fields $K_{j}$ generate the rotation group $SO(3)$ acting on $\M$ confirming that the spacetime metric is spatially isotropic. Spatial homogeneity of any cosmological model follows providing all scalars including components of tensors in the preferred basis $\{X_{a}\}$ or dual $g$-orthonormal cobasis $\{e^{a}\}$ depend only upon the dimensionless evolution variable $\eta$ or upon  real-valued functions of $\eta$. \\

Many observable astrophysical elements are conventionally specified in terms of ``cosmic time''. We denote this by the dimensionless variable $t$ and relate it to $\eta$ by the local diffeomorphism $\wh{\eta}:t\mapsto\eta=\wh{\eta}(t)$ where
\begin{eqnarray}\label{eta_map}
	\frac{d\wh{\eta}(t)}{dt} &=& \frac{1}{\ttype{S}} \qquad\text{and}\qquad \ttype{S} \,=\, (\,\wt{S} \circ \wh{\eta}\,)(t) \,=\, \wt{S}(\,\wh{\eta}(t)\,)\,=\, \etatype{S}.
\end{eqnarray}
It is traditional to fix the map $\wh{\eta}$ so that 
\begin{eqnarray}\label{eta_int}
	\wh{\eta}(t) &=& \int_{0}^{t} \, \frac{dt'}{S(t')}.
\end{eqnarray}
The chain rule is used to relate higher order derivatives of $\etatype{S}$ to derivatives of $S(t)$. Henceforth we distinguish $\eta$-derivatives using Lagrange's `prime' notation and $t$-derivatives by Newton's `dot' notation. By slight abuse of notation, we now write  elements in the cobasis in terms of the evolution variable $t$ as
\begin{eqnarray}\label{CobasisAnst}
	e^{0}\,=\, L_{0}\,dt \qquad\text{and}\qquad e^{k} &=& i\,L_{0}\,\ttype{S}\, \Theta^{k}(r,\theta,\phi) \qquad (k=1,2,3)
\end{eqnarray}
and 
\begin{eqnarray}\label{POLARMETRIC}
	 \g &=& L_{0}^{2}\,\[ -dt \tensor dt + \frac{\ttypesq{S}}{(1+\frac{r^{2}}{4})^{2}}\( dr\tensor dr + r^{2}d\theta\tensor d\theta + r^{2}\sin^{2}(\theta)d\phi\tensor d\phi\)\].
\end{eqnarray}
We note that singularities of the geometry are likely to occur at any zeroes of $S$ (where the cobase  (\ref{CobasisAnst}) collapses). 

\section{The Cosmological Vector-Scalar Model}
\subsection{The Cosmological Vector-Scalar Ans\"{a}tz}
In terms of the preferred spacetime cobasis (\ref{eta_cobas}) induced from the Maurer-Cartan basis $\{i\Theta^{k}\}$  on  $S^{3}$, we adopt the ans\"{a}tz: 
\begin{eqnarray}\label{ProcaAns}
	\ProcaA &=& \ProcaC\, e^{1}
\end{eqnarray}
for the 1-form field where $\ProcaC\in\real$ is dimensionless and 
\begin{eqnarray*}
	\ScalarField \,=\, \etatype{\ScalarField}
\end{eqnarray*}
for the scalar field. Thus in addition to a preferred time-like field $X_{0}$ describing the global fluid flow the model contains the preferred space-like field  $X_{1}$ describing the vector field polarisation  $e^{1}$.
The field equation (\ref{proca_eqn}) then takes the form:
\begin{eqnarray}\label{Spprime}
	\etatypepp{S} &=& -\etatypecube{S}\,\zeta_{0}\,\Upot(\wt{\ScalarField}) - 4\etatype{S}
\end{eqnarray}
and (\ref{psi_eqn}) becomes
\begin{eqnarray}\label{Psipprime}
	\etatypepp{\ScalarField} &=& -\frac{1}{2}\etatypesq{S}\;\zeta_{0}\;\ProcaC^{2}\;\Upot'\hspace{-0.03cm}(\wt{\ScalarField}) - 2\,\etatypep{\ScalarField}\;\frac{\etatypep{S}}{\etatype{S}}
\end{eqnarray}
All elements in these equations are real and dimensionless, including the  constant $\ProcaC$. Once a potential $\Upot$ is adopted for $\etatype{\ScalarField}$ and $\ProcaC,\zeta_{0}$ {\it are assigned non-zero values}, (\ref{Spprime}) and (\ref{Psipprime}) constitute a system of non-linear second-order ODEs for $\etatype{S}$ and $\etatype{\ScalarField}$ amenable to analysis given initial data. Clearly a full analysis of the above coupled system cannot proceed without a specification of the potential $ \Upot $. However since we are interested in particular solutions  with interesting physical implications, one may proceed by assuming only that the potential $\Upot$ has at least one minimum as a function of $\wt{\ScalarField}$, i.e.    $\Upot'\hspace{-0.03cm}(\ScalarField_{0})=0$ and   $\Upot''\hspace{-0.03cm}(\ScalarField_{0})>0$ for some real value $\ScalarField_0$. Then  we observe that the constant field $\etatype{\ScalarField}=\ScalarField_{0}$ is a particular solution to (\ref{Psipprime}) and (\ref{Spprime}) becomes
\begin{eqnarray}\label{Spprime2}
	\etatypepp{S} &=& -\etatypecube{S}\,W - 4\etatype{S}
\end{eqnarray}
where 
\begin{eqnarray*}
	W\equiv\zeta_{0} \,\Xi_{0} \qquad   \text{and} \qquad \Xi_{0}\equiv\Upot(  \ScalarField_{0}).
\end{eqnarray*}
Furthermore, the terms involving $\Gamma_{1}$ in the field equations do not contribute to the cosmological sector. The general solution to (\ref{Spprime2}) can be written in terms of the Jacobi elliptic sine function $\text{sn}$ as:
\begin{eqnarray*}
	\etatype{S} \,=\, \frac{2\sqrt{2\,}\mathcal{K}_{2}}{\mathcal{P}_{1}}\;\;\text{sn}\!\( \frac{(\sqrt{2\lambda_0\,}\,\eta + 2\mathcal{K}_{1})\sqrt{2\,}}{\mathcal{P}_{1}},\, \frac{\mathcal{P}_{0}}{\sqrt{\lambda_0\,}} \)
\end{eqnarray*}
where
\begin{eqnarray*}
	\mathcal{P}_{j} &\equiv& \sqrt{(8-\lambda_0)\mathcal{K}_{2}^{2}+j\lambda_0\;} \qquad\text{and}\qquad \lambda_0\,\equiv\,W+8
\end{eqnarray*}
with arbitrary real constants $\mathcal{K}_{1},\mathcal{K}_{2}$. This {\it general solution}  yields a  potential class of   {\it  physically relevant spacetime cosmographies}  for an analysis of Einstein's equations. Since the  metric scale factor must be real we can simplify this analysis by first transforming to a new evolution variable. \\

We proceed by transforming  our fields to depend on the cosmic time evolution variable $t$ using (\ref{eta_map}) with the composition rules:
\begin{eqnarray*}
	\etatypep{f}		&\equiv& \wt{f}'\hspace{-0.03cm}(\,\wh{\eta}(t)\,) \,=\, \ttypedot{f}\,\ttype{S} \\[0.2cm] 
	\etatypepp{f} 	&\equiv& \wt{f}''\hspace{-0.03cm}(\,\wh{\eta}(t)\,) \,=\, \ttypesq{S}\,\ttypeddot{f} + \ttype{S}\ttypedot{S}\ttypedot{f} 
\end{eqnarray*}
relating the functions $f,\wt{f}$ of one variable. Then (\ref{Spprime}),(\ref{Psipprime}) become for $\ttype{S}\neq 0$:
\begin{eqnarray}
	\label{Sddot}	\ttypeddot{S} &=& -\ttype{S}\,\zeta_{0}\,\Upot(\,\ScalarField(t)\,) - \frac{\ttypedotsq{S}}{\ttype{S}} - \frac{4}{\ttype{S}} \\[0.3cm]
	\nonumber \ttypeddot{\ScalarField} &=& -\frac{1}{2}\zeta_{0}\;\ProcaC^{2}\;\Upot'\hspace{-0.03cm}(\,\ScalarField(t)\,) - 3\,\ttypedot{\ScalarField}\;\frac{\ttypedot{S}}{\ttype{S}}.
\end{eqnarray}
By choosing $\ScalarField_{0}$ as the particular  real root of $\Upot'\hspace{-0.03cm}(\ScalarField_{0})=0$, (\ref{Sddot}) becomes
\begin{eqnarray}\label{SddotU0}
	\ttypeddot{S} &=& -\ttype{S}\,W - \frac{\ttypedotsq{S}}{\ttype{S}} - \frac{4}{\ttype{S}} . 
\end{eqnarray}
 We note that if $S(t)$ is a real solution  to this equation then so is $-S(t)$. Furthermore, for any solution $\ttype{S}$, the expression $a_{0}\ttype{S}$ is {\it not} a solution unless the constant $a_{0}=\pm 1$. Solutions to the  {\it non-linear real}  equation (\ref{SddotU0})  can be constructed  in terms of  complex solutions of an  {\it  auxiliary equation}  for $\ttype{\Sigma}$.  This takes the form
\begin{eqnarray}\label{SigmaDEQ}
	\ttypeddot{\Sigma} &=& - 2\,W\,\ttype{\Sigma} - 8
\end{eqnarray}
with general solution
\begin{eqnarray}\label{SigmaSol}
	\ttype{\Sigma} &=& C_{s1}\cos\!\(\,t\sqrt{\,2\, W\;}\,\) + C_{s2}\sin\!\(\,t\sqrt{\,2\, W\;}\,\) - \frac{4}{\, W}.
\end{eqnarray}
The  constants $C_{s1},C_{s2}$ can be chosen to be real or complex but those  of relevance must ensure that
\begin{eqnarray}\label{St}
	\ttype{S}=\pm\sqrt{\,\ttype{\Sigma}\;}
\end{eqnarray}
takes real values for $t\in\Rint$. From the properties of the hyperbolic functions it is notable that for {\it arbitrary values } of $C_{s1},C_{s2}$ and $W <  0$: 
\begin{eqnarray*}
	\lim_{t\rightarrow\infty}\,\Hub{} \,=\,\sqrt{-  \frac{ \, W} { 2 }        }	
\end{eqnarray*}
in terms of the {\it dimensionless} cosmic time Hubble function\footnote{ This is related to the  {\it observed} cosmic time Hubble function   ${\cal{H}}_{\text{obs}}$ by the relation  ${\cal{H}}_{\text{obs}} =c\mathcal{H}/L_{0}$.      }
\begin{eqnarray*}
	\Hub{} &=& \frac{\ttypedot{S}}{\ttype{S}}.
\end{eqnarray*}
Furthermore it is evident from the properties of the trigonometric functions that solutions with $\, W>0$ are necessary to construct spacetime metrics with a bounded  periodic scale factor $S(t)$ (having many real zeroes) whilst those with $\, W<0$ may generate real solutions with either unbounded  or bounded non-periodic scale factors (with either zero, one or two real zeroes). \\

\SCALEFACVARPARAMS{0.6}{%
	Typical solutions for the scale factor $S(t)$ satisfying (\ref{SddotU0}).
	}
	
Figure~\ref{fig:scalefacvarparams} displays the profiles of $S(t)=\sqrt{\Sigma(t)}$  that have none, one, two or many real zeroes,  obtained by assigning {\it particular} values to $ \{ C_{s1},C_{s2} \} $ and $  \, W  $.  Real solutions $S(t)$ with real zeroes indicate the possibility of geometric spacetime singularities. We conclude that solutions of (\ref{SddotU0})  generate possible ``big-bang-big-crunch''  cyclic cosmological  spacetimes, ``symmetric non-singular'' eternal  cosmological  spacetimes or  ``single big-bang eternally expanding''   cosmological  spacetimes, all with closed spatial universes having the topology of the $3$-sphere.\\

Since the spacetime curvature scalar, $\CS=\text{tr}_{g}(\text{Ric})$, is given in terms of $\ttype{S}$ by
\begin{eqnarray*}
	\ttype{\CS} &=& -\frac{18}{\ttypesq{S}\,L_{0}^{2}} + \frac{6\, W}{L_{0}^{2}},
\end{eqnarray*}
solutions where $\ttype{S}$ is constant yield spacetime geometries with a constant curvature scalar.  An indication of the presence of   singularities in the {\it geometry} (i.e. those that cannot be eliminated by a change of coordinates) is given where $\ttype{\CS}$ becomes unbounded. However, being  bounded  is not a  sufficient condition for the absence of geometric singularities. The Kretschmann scalar $\Kretschmann=\star(R_{ab} \w \star R^{ab})$ offers another invariant that can  also locate geometric singularities where it becomes unbounded. In terms of the scale factor of the model under consideration, it is given by
\begin{eqnarray*}
	L_{0}^{4}\,\ttype{\Kretschmann} \,=\, -6\,W^{2} - \frac{48\, W}{\ttypesq{S}} - 12\,\Hub{2}\, W - \frac{102}{\ttypefour{S}} - \frac{60\Hub{2}}{\ttypesq{S}} - 12\Hub{4}. 
\end{eqnarray*} 
Recent measurements of type-1a supernova luminosities have also made possible observations of the current value of the dimensionless de-acceleration parameter \cite{velten_supernovae,alsabti_handbook}. In terms of cosmic time, this is defined as
\begin{eqnarray*}
	\ttype{q} &=& -\frac{\ttype{S}\,\ttypeddot{S}}{\ttypedotsq{S}}.
\end{eqnarray*}
Using (\ref{SddotU0}) to eliminate the second derivative of the scale factor, this takes the form
\begin{eqnarray}\label{qt1}
	\ttype{q} &=& 1 + \frac{4}{\ttypedotsq{S}} + \frac{\, W}{\Hub{2}}
\end{eqnarray}
which in our model with $\, W<0$, is clearly not manifestly positive  and therefore offers the possibility of finding an epoch $t$  in $\Rint$ where $q(t)<0$
 thereby exploring earlier epochs by retro-diction. 
For  {\it arbitrary values } of $C_{s1},C_{s2}$  in the solutions (\ref{SigmaDEQ}) and when $W <  0$ one has the asymptotic behaviour
\begin{eqnarray*}
	q(t) \,\simeq\, -1 + O(e^{-4Wt})
\end{eqnarray*}
as $t$ approaches infinity. Hence
\begin{eqnarray*} 
	\lim_{t\rightarrow\infty}\,q(t) \,=\, -1
\end{eqnarray*}
reminiscent of de Sitter-type solutions. The recent claim that an accelerating phase  (i.e. $q(t) < 0$) of the cosmos is consistent with observation has recently led to a number of new cosmological models beyond the standard paradigm.\\

\subsection{The Cosmological Maxwell Sector}
We turn next to the construction of  Maxwell fields {\it in the model with} $\ScalarField$ {\it constant} given by $\Upot'(\ScalarField_{0}  )=0,\, \Upot''(\ScalarField_{0}  ) >0$. Since the $U(1)$-invariant Maxwell 2-form $\MaxF=d\MaxA$ must then satisfy the {\it  linear}  equation $\delta\MaxF=\delta\,d\,\MaxA=0$ on $\M$ where $\delta\equiv\star d\star$ is the divergence operator on $p$-forms, a general solution could include a {\it superposition}  of singular solutions and non-singular  modes generated from {\it scalar} harmonics on $S^{3}$. Typical regular modes were considered earlier by Schr\"{o}dinger \cite{bass_photonmass} and have subsequently been developed further in the literature \cite{achour_sphharm,lindblom_harmonics,alertz_EMRW}.\\

Singular solutions would satisfy $\delta\MaxF=\delta\,d\,\MaxA=0$  on electromagnetic source-free domains of $\M$ free of singularities. In Minkowski spacetime certain distributional Maxwell field solutions  serve as models of point particles with electromagnetic multi-pole moments that can interact with regular Maxwell fields. For example  stationary electrically charged point particles are modelled by the Coulomb solution where their charge arises as a de-Rham period  rather than from a 3-dimensional volume integral of any regular charge density. There exist analogues of such solutions in the spacetime with the metric (\ref{gt}).
One readily verifies that in the local chart with coordinates $\{ \eta, r, \theta, \phi \}$ the dimensionless Maxwell 2-form:
\begin{eqnarray*}
	F^I = \frac{Q_0} {4 \pi}\, \( \frac{1}{4} + \frac{1}{r^2}  \) \, dr \w d\eta
\end{eqnarray*}
satisfies $dF^I=\delta F^I=0$ where  $Q_0$ is a dimensionless constant and $r\neq0$. In the preferred frame defined by the unit time-like vector field 
\begin{eqnarray*}
	V \,\equiv\, X_{0} \,=\, \frac{ 1} {L_0 S(\eta) } \frac{\partial}{\partial\eta},
\end{eqnarray*}
this corresponds to an electromagnetic field with electric and magnetic 1-forms:
\begin{eqnarray*}
	{\mathcal  E}^I_{V } &=& i_{V} F^I \,=\, -\frac{Q_0}{4 \pi}\, \( \frac{1}{4} + \frac{1}{r^2}  \)\,\frac{dr}{\etatype{S} L} \,=\, \frac{ Q_0} { 4\pi} \,\(\frac{r}{4}  + \frac{1} {r} \)^{2}\,\frac{ {\mathcal N}} { \etatypesq{S} L^2_0} \\[0.1cm]
	{\mathcal  B}^I_{V } &=&  i_V \star F^I=0
\end{eqnarray*}
in terms of the space-like unit 1-form 
\begin{eqnarray*}
	{\mathcal N} \,=\, \sin(\theta)\cos(\phi)e^{1} + \sin(\theta)\sin(\phi)e^{2} - \cos(\theta)e^{3}.
\end{eqnarray*}	
At any value of $\eta=\eta_0$ the net  electric flux  crossing any 2-sphere $S^2$  surrounding the singularity at $r=0$ is given by the de-Rham period:
\begin{eqnarray*}
	\int_{S^2} \,\star F^I \,=\, Q_0.
\end{eqnarray*}
One notes that the norms of $F^I$ and $ {\mathcal  E}^I_{V}$ coincide and:
\begin{eqnarray*}
	|F^{I}| \,=\, \sqrt{\star\(F^{I} \w \star F^{I}\)\,} \,=\, \frac{Q_0} {16\pi\etatypesq{S} L^2_0} \,\(\frac{r}{4}  + \frac{1}{r} \)^{2}
\end{eqnarray*}
implying that both $|F^{I}|,|\mathcal{E}^{I}_{V}|$ diverge as $r\rightarrow\infty$. However the anti-podal point of $S^3$ where $r=\infty$  is not in the local chart. But  $F^I$, $\g$ and the coframe $\{e^{1},e^{2},e^{3}\}$ are each invariant under the  symmetry transformation $\{ Q_0 \mapsto - Q_0,     r\mapsto \frac{4}{r}$\} implying that  the above local representation of $F^{I}$ {\it extends} to a global electrically neutral distributional Maxwell solution with singularities at a pair of antipodal points of $S^3$ with equal and opposite  de-Rham periods  at each pole (see appendix). Since  this  zero-total electric charge solution on $S^3$ has a {\it pair} of spatially distinct singularities we shall designate it an {\it electric bi-pole} to distinguish it from  electrically neutral singular {\it dipole} solutions that each possess a {\it single}  spatially located singularity.\\

Similarly  Hodge-duality implies the existence of  distributional  {\it magnetic bi-pole}
solutions  of the form:
\begin{eqnarray*}
	F^{II} \,=\,  \frac{M_0} {4\pi} \, \sin(\theta) \, d\theta\wedge d\phi
\end{eqnarray*}
where $M_{0}$ is a dimensionless constant and
\begin{eqnarray*}
	{\mathcal  E}^{II}_{V } &=& i_{V}F^{II}\,=\, 0 \\[0.1cm]
	{\mathcal  B}^{II}_{V } &=& i_V \star F^{II} \,=\, \frac{M_0} {4\pi}\(\frac{1}{4}+\frac{1}{r^{2}}\)\, \frac{dr}{\etatype{S}L_{0}} \,=\, -\frac{M_0} {4\pi}\(\frac{r}{4}+\frac{1}{r}\)^{2}\, \frac{\mathcal{N}}{\etatypesq{S}L_{0}^{2}}
\end{eqnarray*}
in the chart with coordinates $\{ \eta, r, \theta, \phi \}$,  with constant dimensionless {\it magnetic}  charge:
\begin{eqnarray*}
	\int_{S^2} \, F^{II} \,=\, M_0.
\end{eqnarray*}
Thus  $F^{II}$  represents a global magnetically  neutral distributional Maxwell solution with singularities at a pair of antipodal points of $S^3$ with equal and opposite  de-Rham periods at each pole and symmetric under the
transformation $\{ M_0 \mapsto - M_0,     r\mapsto \frac{4}{r}$\}.
Despite the similarity of  these Maxwell field LFRW bi-pole solutions to the  Minkowski Coulomb and magnetic monopole solutions in the vicinity of their isolated singularities, neither $F^I$ nor $F^{II}$  is a  spatially spherically symmetric solution:
\begin{eqnarray*}
	{\mathcal L }_{K_j}  {F}^{I} \neq 0
	\quad\text {   and  }   \quad
	{\mathcal L }_{K_j}  {F}^{II} \neq 0\qquad (j=1,2,3)
\end{eqnarray*}
in terms of the spacelike Killing vectors $\{K_{j}\}$ on $\M$ discussed in section~3.\\

In the cosmological context that follows we shall content ourselves with a particular {\it regular source-free superposition}  that displays interesting  distinct electromagnetic features  and consider the particular
 ans\"{a}tz:
\begin{eqnarray*}
	\MaxF &=& d\(\,\etatype{\alpha}\,e^{1}\,\) + \etatype{E}\,e^{0}\w e^{1} +  \etatype{B}\,e^{2}\w e^{3}
\end{eqnarray*}
in terms of the  real scalar functions $ \widetilde\alpha, \widetilde E, \widetilde B $ of $\eta$ and return to a discussion of local charge fluctuations in section 6. Now the source free Maxwell equations  $d\MaxF=\delta\MaxF=0$ lead to the  non-distributional  solutions:
\begin{eqnarray*}
	 \etatype{E} &=& \frac{\[C_{k1}\cos(2\eta) - C_{k2}\sin(2\eta)\]}{\etatypesq{S}},\qquad \etatype{B} \;=\; \frac{\[C_{k1}\sin(2\eta) + C_{k2}\cos(2\eta)\]}{\etatypesq{S}} 
\end{eqnarray*}
and the second-order ODE:
\begin{eqnarray*}
	\etatypepp{\alpha} - W\;  {\etatypesq{S}}\,\etatype{\alpha} + 2\etatypep{\alpha}\;\frac{\etatypep{S}}{\etatype{S}} &=& 0
\end{eqnarray*}
for $\etatype{\alpha}$. In terms of the cosmic time variable $t$:
\begin{eqnarray}\label{MAXWELLF_t}
	\MaxF &=& d\(\,\ttype{\alpha}\,e^{1}\,\) + \ttype{E}\,e^{0}\w e^{1} +  \ttype{B}\,e^{2}\w e^{3}
\end{eqnarray}
where
\begin{eqnarray*}
	 \ttype{E} &=& \frac{\[C_{k1}\cos(\,2\,\wh{\eta}(t)\,) - C_{k2}\sin(\,2\,\wh{\eta}(t)\,)\]}{\ttypesq{S}},\quad \ttype{B} \,=\, \frac{\[C_{k1}\sin(\,2\,\wh{\eta}(t)\,) + C_{k2}\cos(\,2\,\wh{\eta}(t)\,)\]}{\ttypesq{S}} 
\end{eqnarray*}
and $\ttype{\alpha}$ is a solution of
\begin{eqnarray}\label{MaxAlphaDEQt}
	\ttypeddot{\alpha} + 3\,\Hub{}\;\ttypedot{\alpha} - W \ttype{\alpha} &=& 0.
\end{eqnarray}
This is a linear non-autonomous ordinary differential equation for $\ttype{\alpha}$ which can be readily integrated in terms of two arbitrary real constants. Depending upon the sign of $\, W$, it resembles the dynamics of an oscillator or an ``anti-oscillator'' with time-dependent damping or ``anti-damping'' determined by the  dimensionless Hubble function $\Hub{}$. \\

The particular solutions involving $\etatype{E}$ and $\etatype{B}$  above generalize the source-free Maxwell solutions found by Kopi\'{n}ksi and Nat\'{a}rio \cite{kopinski_GRG} in the context of a standard LFRW model involving two non-interacting null fluids. To transform these to a cosmic time variable requires the computation of the  integral (\ref{eta_int}).  To simplify our analytic presentation in terms of the evolution variable $t$,  in the following we shall therefore generate a regular Maxwell field from the restricted Maxwell potential 1-form ans\"{a}tz $\MaxA=\ttype{\alpha}\,e^{1}$ by analogy with the ans\"{a}tz $\ProcaA=\ProcaC\,e^{1}$. \\

When $\etatypeZ{S}\neq 0$   and $  \etatypeZ{\alpha}  $ is regular one can immediately calculate the {\it instantaneous} Maxwell magnetic helicity 3-form $\widetilde\MagHelM\!\left.\right|_{\eta=\eta_{0}}\equiv \MaxA \w \MaxF|_{\Sigma_{3}}$ on any 3-dimensional space-like hypersurface $\Sigma_{3}$ defined by $\eta=\eta_{0}\in\real$,  in terms of $ \etatypeZ{S} $  and   $\etatypeZ{\alpha}$:
\begin{eqnarray*}
	\widetilde{\MagHelM}\!\left.\right|_{\eta=\eta_{0}} &=& \left.-2\frac{\etatypesqZ{\alpha}}{L_{0}\,\etatypeZ{S}}\, e^{1} \w e^{2} \w e^{3}\right|_{\Sigma_{3}}
\end{eqnarray*}
where  $\Sigma_{3}$ is, topologically,  the 3-sphere $S^3$. This  gives a {\it finite} total instantaneous Maxwell magnetic helicity:
\begin{eqnarray*}
	\int_{S^{3}}\widetilde\MagHelM\!\left.\right|_{\eta=\eta_{0}} &=& -2\frac{\etatypesqZ{\alpha}}{L_{0}\,\etatypeZ{S}}\, \Vol(S^{3}) \,=\, - 4\pi^{2}\,\etatypesqZ{\alpha}\,L_{0}^{2}\,\etatypesqZ{S} 
\end{eqnarray*}
since $\displaystyle\Vol(S^{3})=\int_{S^{3}}\left.e^{1}\w e^{2} \w e^{3}\right|_{\Sigma_{3}}=2\pi^{2}\,L_{0}^{3}\,\etatypecubeZ{S}$.\\

In a similar way, we may define the {\it instantaneous} $\ProcaA$-magnetic helicity associated with the vector field configuration $\etatype{\ProcaA}$ in (\ref{ProcaAns}): 
 $\widetilde\MagHelP\!\left.\right|_{\eta=\eta_{0}}\equiv \ProcaA \w d\ProcaA|_{\Sigma_{3}}$ and derive, for $\etatypeZ{S}\neq 0$,  the finite total instantaneous $\ProcaA$-magnetic helicity:
\begin{eqnarray*}
	\int_{S^{3}}\widetilde \MagHelP\!\left.\right|_{\eta=\eta_{0}} &=& -2\frac{\ProcaC^{2}}{L_{0}\,\etatypeZ{S}}\, \Vol(S^{3}) \,=\, - 4\pi^{2}\,\ProcaC^{2}\,L_{0}^{2}\,\etatypesqZ{S} .
\end{eqnarray*}
The existence of non-zero total instantaneous magnetic helicity for the particular solutions where $\ProcaA$ and $\MaxA$ are proportional to the 1-form $e^{1}$ on each constant $\eta$ hypersurface implies that the corresponding electric and magnetic field lines (i.e. the instantaneous integral curves of the electric and magnetic field vectors),  defined by splitting the 2-form field strengths $\MaxF$, $\ProcaF=d\ProcaA$ respectively into electric and magnetic type 1-form fields relative to the preferred coframe (\ref{TLUV}), are ``linked'' and exhibit a ``Hopfian'' knot configuration at each instant $\eta=\eta_{0}$ \cite{berger_maghel,irvine_linkedlight,ranada_maghel}. \\

It is worth noting that the  Maxwell gauge field $\MaxA=\etatype{\alpha}\,e^{1}$ satisfies the condition $\delta\MaxA=0$. This reduces its number of independent  physical degrees of freedom to two. Although in general the 1-form $\ProcaA$ is not a $U(1)$ gauge field and does not satisfy $\delta\ProcaA=0$, when one restricts to a model where the constant  scalar field satisfies $\Upot'\hspace{-0.03cm}(\ScalarField_{0})=0$, $\ScalarField_{0}\in\real$, then $\delta\ProcaA=0$ also. \\

\subsection{The Cosmological Einstein Sector}
Next we turn attention to the total \SEM $T$ in the Einstein equation (\ref{OurEFE}). Introducing the dimensionless fluid density $\ttype{\Omega}=\ttype{\rho}\,L_{0}^{2}c^{2}/\kappa$ and dimensionless fluid pressures $\ttype{P_{k}}=\ttype{p_{k}}\,L_{0}^{2}/\kappa,$ and substituting the equations $\ScalarField=\ScalarField_{0}$, $\MaxA=\ttype{\alpha}\,e^{1}$, $\ProcaA=\ProcaC\,e^{1}$ into $T$ yields the non-zero $\g$-orthonormal  $T$ components:
\begin{eqnarray*}
	\ttype{T_{00}} &=& -\mathcal{J}^{-}(t) + \frac{2\ttypesq{\alpha}}{\ttypesq{S}}  + \frac{\Hub{2}\ttypesq{\alpha}}{2} + \Omega(t) \\[0.2cm]
	\ttype{T_{11}} &=& \mathcal{J}^{+}(t)  - \frac{2\ttypesq{\alpha}}{\ttypesq{S}} - \frac{\Hub{2}\ttypesq{\alpha}}{2} - 2\,W+ P_{1}(t)\\[0.2cm]
	\ttype{T_{22}} &=& -\mathcal{J}^{+}(t)   + \frac{2\ttypesq{\alpha}}{\ttypesq{S}}  + \frac{ \Hub{2}\ttypesq{\alpha}}{2} - 2\,W+ P_{2}(t)\\[0.2cm]
	\ttype{T_{33}} &=& -\mathcal{J}^{+}(t)  + \frac{2\ttypesq{\alpha}}{\ttypesq{S}}  + \frac{ \Hub{2}\ttypesq{\alpha}}{2} - 2\,W+  P_{3}(t)
\end{eqnarray*}
in terms of the dimensionless Hubble function $\Hub{}$ and where 
\begin{eqnarray*}
	\mathcal{J}^{\pm}(t) &\equiv& \(\pm \frac{\, W}{2} - \frac{2}{\ttypesq{S}} - \frac{\ttypedotsq{S}}{2\ttypesq{S}} \)\ProcaC^{2} - \frac{\ttypedotsq{\alpha}}{2}\( \frac{}{}1+2\,\ttype{\alpha}\Hub{}\,\).
\end{eqnarray*}
i.e.
\begin{eqnarray*}
	\mathcal{J}^{\pm}(t) &\equiv& \(\pm \frac{\, W}{2} - \frac{2}{\ttypesq{S}} -  \frac{1}{2}  \Hub{2}    \)\ProcaC^{2} - \frac{\ttypedotsq{\alpha}}{2}\( \frac{}{}1+2\,\ttype{\alpha}\Hub{}\,\).
\end{eqnarray*}
It remains to solve the dynamical system containing  (\ref{MaxAlphaDEQt}) and 
\begin{eqnarray}\label{EIN_EQN_t}
	\kappa\,\EIN_{t} &=& T_{t} \,=\, \TT_{t} + \mathbb{T}_{t}
\end{eqnarray}
where $\{\EIN_{t},T_{t},\TT_{t},\mathbb{T}_{t}\}$ is the form taken by the tensors $\{\EIN,T,\TT,\mathbb{T}\}$ with the ans\"{a}tz (\ref{CobasisAnst}), (\ref{ProcaAns}), (\ref{MAXWELLF_t}) and $\ScalarField=\ScalarField_{0}$. This system comprises six ordinary coupled equations for the six variables: 
\begin{eqnarray*}
	\{\, \,\ttype{\Omega}, \,\ttype{P_{1}}, \,\ttype{P_{2}}, \,\ttype{P_{3}}, \,\ttype{S}, \,\ttype{\alpha}\,\}.
\end{eqnarray*}
However, any solutions will automatically satisfy the Bianchi identity $\nabla\cdot\EIN=0$ which implies $\nabla\cdot T=0$. Hence there must be some relations between the six solutions. For any solutions $\ttype{\alpha}, \ttype{S}$, the solutions for the fluid variables are found to be
\begin{eqnarray}
\begin{split}\label{EOS1}
	\ttype{\Omega}	&\,=\, \mathcal{J}^{-}(t)  + \frac{(3-2\ttypesq{\alpha})}{\ttypesq{S}}  + \frac{\Hub{2}}{2}\( 6-\ttypesq{\alpha}\) \\[0.2cm]
	\ttype{P_{1}} 	&\,=\, -\mathcal{J}^{+}(t)   + \frac{(7+2\ttypesq{\alpha})}{\ttypesq{S}}  + \frac{\Hub{2}}{2}\( 2+\ttypesq{\alpha}\) + 2\,W\\[0.2cm]
	\ttype{P_{2}} 	&\,=\, \mathcal{J}^{+}(t)  + \frac{(7-2\ttypesq{\alpha})}{\ttypesq{S}}  + \frac{\Hub{2}}{2}\( 2-\ttypesq{\alpha}\) + 2\,W\\[0.2cm]
	\ttype{P_{3}} 	&\,=\, \mathcal{J}^{+}(t)   + \frac{(7-2\ttypesq{\alpha})}{\ttypesq{S}}  + \frac{\Hub{2}}{2}\( 2-\ttypesq{\alpha}\) + 2\, W.
\end{split}
\end{eqnarray}
Two algebraic relations amongst these fluid variables emerge:
\begin{eqnarray*}
	\ttype{\Omega} &=& -\, W\,\ProcaC^{2} + \frac{1}{2}\(\ttype{P_{2}} - \ttype{P_{1}}\,\) + \frac{3}{\ttypesq{S}} + 3\Hub{}^{2}
\end{eqnarray*}	
and
\begin{eqnarray*}
	\ttype{P_{2}} &=& \ttype{P_{3}}.
\end{eqnarray*}
These relations parametrize a {\it dynamically induced}  anisotropic fluid {\it mechanical} equation of state which is, in general, manifestly dependent on the evolution variable $t$ through the dynamics of the scale factor $\ttype{S}$ and the vector, scalar and Maxwell fields. Note that although the fluid \SEM is thereby anisotropic \cite{hazeltine_anisoplasma}, the {\it total} \SEM $T_{t}$ is not. Spatial anisotropies in the field contributions to $T_{t}$ cancel those in the fluid contribution  $\mathbb{T}_{t}$  yielding the isotropic Einstein tensor ${\EIN}_{t}$ with non-zero $\g$-orthonormal components $\EIN_{t}(X_{a},X_{b})=\ttype{\EIN_{ab}}$ given by:
\begin{eqnarray}
\begin{split}\label{EIN_COMPS}
	 \ttype{\EIN_{00}} &\,=\, \frac{3}{L_{0}^{2}\,\ttypesq{S}}\( 1 + \ttypedotsq{S}\)  
	\equiv\frac{3}{L_0^2}\( \frac{1}{ \ttypesq{S} }   + \Hub{2}   \)       
	\\[0.3cm]
	 \ttype{\EIN_{kk}} &\,=\, -\frac{1}{L_{0}^{2}\,\ttypesq{S}}\( 1 + 2\ttype{S}\ttypeddot{S} + \ttypedotsq{S} \)   \equiv  
	\frac{1}{L_0^2}\(   \Hub{2}( 2 q(t)-1 )    - \frac{1}{ \ttypesq{S} }       \)
\end{split}
\end{eqnarray} 
for $k=1,2,3$.

\section{Testing the Cosmological Model}
In the previous section we showed that   the ans\"{a}tz (\ref{ProcaAns})  together with  $\ttype{\ScalarField}=\ScalarField_{0} $ at a minimum  $\Upot_{0}$  of $\Upot$ was sufficient to reduce the coupled scalar-vector system (\ref{psi_eqn}) and (\ref{proca_eqn})  to the single condition (\ref{SddotU0})  to determine the scale function $S(t)$, the general solution depending on  the constants of integration $\{ C_{s1} , C_{s2} \}$ and the parameter 
\begin{eqnarray*}
	W\,\equiv\, \zeta_0\, \Xi_{0}. 
\end{eqnarray*}
Solutions for the fluid variables $\Omega(t), P_1(t),P_2(t),P_3(t)$ then follow from the Einstein equations (\ref{EIN_EQN_t})   in terms of cosmological solutions for the Maxwell equations (\ref{max_eqn}).\\

An immediate test of the model can be made by exploiting currently available {\it measured values}  of the Hubble parameter $\wh{\cal H}_{\text{obs}}(\wh{t}_{0})$ and  the de-acceleration  parameter $\wh{q}_{\text{obs}}(\wh{t}_{0})$ since these are independent of the standard cosmological paradigms based on particular multi-fluid equations of state. The cosmic time evolution variable  $\wh{t}_{0}$ here,  with dimensions of time,  is related to the current dimensionless time coordinate $t_0$ by the relation $\wh{t}_{0}= t_0 L_0/c $.  Furthermore, since $[ \wh {\cal H}  ]  = \text{T}^{-1}$, if one uses SI units of time in seconds, then the parameter $L_0=1\text{m}$ and  our dimensionless  Hubble parameter ${\cal H}  (t_{0})=\wh{\cal H}_{\text{obs}} (\wh{t}_0)/c$. Taking $\wh{\cal H}_{\text{obs}} (\wh{t}_{0})= 58\,h \,\text{km}\, \text{sec}^{-1}/\text{Mpc}\, =3.243 \times 10^{-18}\, h \,\text{sec}^{-1}$ yields ${\cal H}  (t_{0})=1.082 \times 10^{-26}  \,h $ with $\vert h \vert=0.58 $ \cite{weinberg_cosmology} while the dimensionless value of $\wh{q}_{\text{obs}}(\wh{t}_0)=q(t_0)=-0.56$ is compatible with current observation. \\

Since (\ref{qt1}) implies the de-acceleration parameter $q(t)>0$ for $W>0$, contrary to current observation, we will concentrate attention on solutions $S(t)$ of (\ref{SddotU0}) with $W<0$. Particular solutions are then fixed by specifying values for $W,S(0)$ and $\dot S(0)$ since there is no loss of generality by labelling $t_0=0$. The values of $S(0)$ and $\dot S(0)$ are then in turn furnished in terms of values for ${\cal H}(0)$ and $q(0)$ for some $W<0$. Evaluating (\ref{SddotU0}) at $t=0$ yields  the following  value for $S(0)$:
\begin{eqnarray}\label{S0}
	S(0) &=& \frac{ 2 }  { {\cal H}(0)\,\sqrt{ q(0)+\Delta^2-1  }   }	 
\end{eqnarray}
where $W= - {\cal H}(0)^2\,\Delta^2$ with $\Delta \in \real$. Then 
\begin{eqnarray*}
	\dot{S}(0) \,=\,	 {\cal H}(0)\, S(0)
\end{eqnarray*}
and the solution $S(t)$ is fixed and real for any $\Delta^{2}>1-q(0)$.\\

\OBSERVDATAPLOTS{0.95}{%
	Astrophysical  model  predictions based on current $(t=0$)   values of the dimensionless Hubble parameter
	${\cal H}(0)$  and  the de-acceleration parameter $q(0)$. 
	}

Figure~\ref{fig:observdataplots} shows cosmological characteristics based on the solution\footnote{In terms of the complex constants in (\ref{St})  this data corresponds to the values: $C_{s1}=5.002 \,\,\times 10 ^{54} $, $C_{s2}=-5.700 i \,\,\times 10 ^{54}$ and $W=-6.223\,\times 10^{-53}$. } obtained with $\Delta=1.257$, giving $S(0)=2.25 \times 10^{27}$ and retro-dicts the location of a single zero of $S$ at $t=-1.245 \times 10^{26}$
together with a non-uniformly exponentially increasing value of $S(t)$. The curves for $ {\mathcal K}    $  and  $ {\mathcal R} $ clearly show that the zero of $S$ is a geometric singularity so one may conclude that the model  fitting $  {\cal H}$ and $q$ with $\Delta=1.257$ at the current epoch yields an age of the Universe to be around $13.2 \times 10^9$ years. The second figure displays the past and future behaviour of the de-acceleration function $q(t)$ while the fifth figure displays the past and future behaviour of the dimensionless Hubble  function ${\cal H}(t)$. Although there is no data for the current value of the dimensionless ``jerk'' function:
\begin{eqnarray*}
	j(t) \,=\,	\frac{S(t)^2\, \dddot S(t)  }{   \dot S(t)^3  },
\end{eqnarray*}
the model predicts $j(0)=1.480$ and the last plot in  Figure~\ref{fig:observdataplots} displays its past and future behaviour. \\

In general, when $W<0$ one has:
\begin{eqnarray*} 
	\lim_{t\rightarrow\infty}\,j(t) \,=\, 1 	
\end{eqnarray*}
for {\it arbitrary values } of $C_{s1},C_{s2}$  in the solutions  of (\ref{SigmaDEQ}).\\

Although all these cosmographic features are mathematically independent of the Einstein equations \cite{visser_cosmography}, they are consistent with a  single ``big-bang'' scenario undergoing eternal  non-uniform exponential  accelerated expansion. The Einstein equations in turn then predict the behaviour of the density and pressures in the fluid stress-energy-momentum tensor compatible with the Maxwell, scalar and vector field  stress-energy-momentum tensors,  determined from their associated field equations. It is worth noting at this stage that none of the particular solutions that follow from the equations for $\ttype{\alpha}, \ttype{S}, \ttype{\ScalarField}, {\ProcaA}$, give rise to a term in $T$ of the form $\Lambda_{0}\,\g$ with $\Lambda_{0}$ constant and $\g$ the spacetime metric tensor (i.e. to an effective cosmological {\it constant}).\\

\ENERGYCONDS{Plots displaying constraints for the Dominant, Strong, Weak and Null Energy Conditions respectively. Some curves are indistinguishable in the plots.}

It is also of interest to examine what the above cosmological solution implies for the various  local ``energy conditions'' that feature in a number of theoretical criteria for general relativistic stress-energy-momentum tensors. These are based on inequalities that involve the eigenvalues of such symmetric tensors.  Thus one may express the right hand side of the Einstein equation  (\ref{EIN_EQN_t}) as
\begin{eqnarray*}
	{T}_{t} &=& \frac{\ttype{\wh{\Omega}}}{L_{0}^{2}}\,e^{0} \tensor e^{0} + \frac{1}{L_{0}^{2}}\sum_{k=1}^{3}\ttype{\wh{P}_{k}}\,e^{k}\tensor e^{k}.
\end{eqnarray*}
in terms of the  {\it eigen-mass-energy density} $\ttype{\wh{\Omega}}$ and {\it eigen-pressures} $\ttype{\wh{P}_{k}}$ and note that the total mass-energy of the interacting system includes the mass-energy of the fields as well as the material  fluid. The  eigenvalues of $T_{t}$ following simply from the eigenvalues of  ${\EIN}_{t}$ displayed in (\ref{EIN_COMPS}). Cosmological stress-energy-momentum tensors that satisfy these local ``energy conditions'' at cosmic time $t$ obey the constraints on the left in figure~\ref{fig:energyconds} and the figures on the right illustrate the behaviour of the expressions in the Dominant, Strong, Weak and Null energy inequalities as a function of dimensionless cosmic time over an interval, starting from the big-bang and including  the current epoch ($t=0$).  Over this interval all inequalities are satisfied except the Strong Energy Condition. Opinion seems divided on the significance of this condition in constraining cosmological models \cite{visser_energyconds}.

\section{A Scalar and Vector Field Fluctuation Model}
With a dynamic model  that yields an evolving scale factor one may construct a recursive  framework for analysing  the {\it fluctuation and thermal history}  of the  matter and {\it field}  content of the Universe  consistent with the solutions (\ref{EOS1}) to  Einstein's equations (\ref{OurEFE}) without a cosmological constant. This  may offer some insight into   the nature of ``dark-matter'' \cite{daly_model}. \\

It has been shown in section 4 that the ans\"{a}tz  (\ref{CobasisAnst}), defining the metric $g$,  and  the ans\"{a}tz (\ref{ProcaAns})  for $\ProcaA$ yield analytic solutions to the  equations  (\ref{psi_eqn}), (\ref{proca_eqn}) provided the constant solution $\ScalarField=\ScalarField_{0}$ satisfies $\Upot'(\ScalarField_{0})\,=\, 0$ and the metric scale factor $S(t)$ is a solution of the non-linear ODE  (\ref{SddotU0}), with the general solution given by (\ref{St}). Any particular solution of (\ref{MaxAlphaDEQt}) then generates a  source free Maxwell field that completes a cosmological solution to the coupled field system   (\ref{psi_eqn}), (\ref{max_eqn}),  (\ref{proca_eqn})  compatible with Einstein's equations.\\

In section 5 particular solutions $S(t)$  to  (\ref{SddotU0}) have been shown to describe  distinct cosmological spacetimes including those  possessing a {\it single}  initial singularity in the spacetime geometry. However, given $\Upot(\ScalarField)$, the fully coupled {\it general system} (\ref{psi_eqn}), (\ref{max_eqn}), (\ref{proca_eqn}), (\ref{OurEFE})  has other solutions that can break the spatial homogeneity and isotropy of the particular LFRW metric solution. Such solutions may, given additional physical data, ultimately account for the formation of plasma states in the early Universe, needed  to  thermalise matter and electromagnetic radiation, and  for the formation of the elements, stars and galaxies during later epochs of its evolution.\\

A tentative approximation scheme, adopted below, is to generate a sequence of linearised perturbations
of the  equations   (\ref{psi_eqn}), (\ref{max_eqn}), (\ref{proca_eqn}) about the exact cosmological solution described in section 5 in terms of the scale factor (\ref{St}) in the LFRW metric. This particular scheme assumes the evolving perturbations have a negligible back-reaction on the ambient LFRW gravitational field. This is reasonable for field fluctuations that are localised in small regions of space. In section 8 we will also describe analytically,  field perturbations  that are {\it spatially  localised}  during epochs were the ambient gravitational field is itself neglected, i.e. where the dominant physics under consideration  takes place in a background Minkowski spacetime.\\

Working to order $\epsilon^{2}$ in the dimensionless  perturbed fields, let:
\begin{eqnarray}\label{PerturbExpansion}
	\begin{split}
		\ScalarField &\,=\, \indexScalarField{0} + \epsilon\indexScalarField{1} + \epsilon^{2}\indexScalarField{2} \\[0.2cm]
		\ProcaA &\,=\, \indexProcaA{0} + \epsilon\indexProcaA{1} + \epsilon^{2}\indexProcaA{2} \\[0.2cm]
		\MaxA &\,=\, \indexMaxA{0} + \epsilon\indexMaxA{1} + \epsilon^{2}\indexMaxA{2}
	\end{split}
\end{eqnarray}
with norms $  \vert \ScalarField^{ (j)} \vert \le 1,   \,\,     \vert \star(\indexProcaA{j}  \wedge  \star    \indexProcaA{j} )  \vert   \le 1,   \,\,      \vert  \star(\indexMaxA{j}  \wedge  \star    \indexMaxA{j} )      \vert   \le 1   \,\,   $ for $j=1,2$. Up to this maximum order it is sufficient to write the potential  $\Upot$ in the form of a  truncated Taylor series about the constant value $\ScalarField_0$ where it has a  {\it local minimum}:
\begin{eqnarray*}
	\Upot(\ScalarField) \,=\, \Xi_{0} + \frac{1}{2}\Xi_{1}(\ScalarField-\ScalarField_{0})^{2} + \frac{1}{6}\Xi_{2}(\ScalarField-\ScalarField_{0})^{3} 
\end{eqnarray*}
with 
\begin{eqnarray*}
     \Xi_{0} \equiv\, \Upot(\ScalarField_{0}),\qquad
	\Upot'(\ScalarField_{0})\,=\, 0, \qquad \Xi_{1} \,\equiv\, \Upot''(\ScalarField_{0})\,>\,0 \qquad\text{and}\qquad  \Xi_{2} \,\equiv\, \Upot'''(\ScalarField_{0}).
\end{eqnarray*}
In terms of the summands in the  expansion of $\ScalarField$ above, it follows that to order $\epsilon^2$:
\begin{eqnarray}\label{Taylor}
	 \Upot(\ScalarField) \,=\, \Xi_{0} + \frac{1}{2}\Xi_{1}(\indexScalarField{1})^{2}\,\epsilon^{2}, \qquad
	\Upot'(\ScalarField) \,=\, \Xi_{1}\indexScalarField{1}\epsilon + \(\Xi_{1}\indexScalarField{2}+\frac{1}{2}\Xi_{2}(\indexScalarField{1})^{2}\)\epsilon^{2}.
\end{eqnarray}
Substituting  the expansions (\ref{PerturbExpansion}), (\ref{Taylor}) into the PDE equations (\ref{psi_eqn})-(\ref{proca_eqn})	 yields the zeroth $\epsilon$-order PDE system:
\begin{eqnarray}
	\nonumber	d\star d\indexScalarField{0} \,=\, 0 \\[0.2cm]
	\nonumber	d\star d\indexProcaA{0} + \frac{W}{L_{0}^{2}}\star \indexProcaA{0} \,=\, 0 \\[0.2cm]
	\label{ZeroOrderMax}	d\indexMaxF{0} \,=\, 0, \qquad d\star\indexMaxF{0} \,=\, 0, 
\end{eqnarray}
the first $\epsilon$-order PDE system:
\begin{eqnarray}
	\label{FirstOrderPsi_1}	d\star d\indexScalarField{1} + \frac{\Xi_{1}}{2}\(\NewCoupling\indexMaxF{0} \w \indexMaxF{0} - \frac{\zeta_{0}}{L_{0}^{2}}\,\indexProcaA{0} \w \star \indexProcaA{0} \)\indexScalarField{1} \,=\, 0 \\[0.2cm]
	\label{FirstOrderProca}	d\star d\indexProcaA{1} + \frac{W}{L_{0}^{2}}\star \indexProcaA{1} \,=\, 0 \\[0.2cm]
	\label{FirstOrderMax}	d\indexMaxF{1} \,=\, 0, \qquad d\star\indexMaxF{1} \,=\, 0 , 
\end{eqnarray}
and the second $\epsilon$-order PDE system:
\begin{eqnarray}
	\label{SecOrderPsi}		d\star d\indexScalarField{2} + \frac{\Xi_{1}}{2}\(\NewCoupling\,\indexMaxF{0} \w \indexMaxF{0} - \frac{\zeta_{0}}{L_{0}^{2}} \,\indexProcaA{0} \w \star\indexProcaA{0}\)\indexScalarField{2} \,=\, J \\[0.2cm]
	\label{SecOrderProca}	d\star d\indexProcaA{2} + \frac{W}{L_{0}^{2}}\star \indexProcaA{2} \,=\, \mathbb{J} \\[0.2cm]
	\label{SecOrderMax}		d\indexMaxF{2} \,=\, 0, \qquad d\star\indexMaxF{2} \,=\, \mathcal{J}  
\end{eqnarray}
where
\begin{eqnarray*}
	\indexMaxF{j}=d\,\indexMaxA{j},  \qquad (j=0,1,2)
\end{eqnarray*}
and
\begin{eqnarray*}
	J &\equiv& + \( \frac{\zeta_{0}\Xi_{1}}{L_{0}^{2}}\,\indexProcaA{0} \w \star \indexProcaA{1}-\frac{\NewCoupling\Xi_{1}}{2}\(\indexMaxF{1} \w \indexMaxF{0} + \indexMaxF{0} \w \indexMaxF{1}\) \)\indexScalarField{1}	\\[0.2cm]
		&& \quad + \( \frac{\zeta_{0}\Xi_{2}}{4L_{0}^{2}}\, \indexProcaA{0} \w \star \indexProcaA{0} - \frac{\NewCoupling\Xi_{2}}{4}\, \indexMaxF{0} \w \indexMaxF{0}\)(\indexScalarField{1})^{2} \\[0.5cm]
	\mathbb{J} &\equiv&  -\frac{\Xi_{1}\zeta_{0}}{2L_{0}^{2}}\,(\indexScalarField{1})^{2}\,\star \indexProcaA{0} \\[0.5cm]
	\mathcal{J} &\equiv& -\NewCoupling\Xi_{1}\,\indexScalarField{1}\,d\indexScalarField{1} \w \indexMaxF{0} \,=\, -\NewCoupling\Xi_{1}\,d\( \frac{1}{2}(\indexScalarField{1})^{2} \w \indexMaxF{0}\).
\end{eqnarray*}
 
One sees from these equations that, given $\indexScalarField{0},  \indexProcaA{0}$ and $\indexMaxF{0}$,
the scheme generates  a sequence of PDE systems for $\indexScalarField{j},  \indexProcaA{j}$ and $\indexMaxF{j}$.
With $ \indexScalarField{0} = \ScalarField_0, \,\,\,\Upot'(\ScalarField_0)=0$ and  $\indexProcaA{0}=\Gamma_1\,e^1$, the zeroth-$\epsilon$ order  system is satisfied by solving the {\it source-free} Maxwell system  (\ref{ZeroOrderMax}). Thus these zeroth $\epsilon$-order  equations are those used in sections 2, 3 and 4 to establish a viable cosmological model when combined with the Einstein equations. The first $\epsilon$-order field modifications  to this exact cosmological solution (\ref{FirstOrderPsi_1})-(\ref{FirstOrderMax}) are given by solutions to an uncoupled, {\it homogeneous}, (source-free)  second-order PDE system. By contrast, at the next $\epsilon$-order, the perturbations (\ref{SecOrderPsi})-(\ref{SecOrderMax}) satisfy uncoupled {\it inhomogeneous} equations with sources generated by  solutions to the lower $\epsilon$-order systems.\\

Of particular note is the perturbed Maxwell system  (\ref{SecOrderMax})	 for the perturbation  $\indexMaxF{2}$ containing the  manifestly conserved electric current 3-form $\mathcal{J}$:
\begin{eqnarray*}
	d\,\mathcal{J}=0.
\end{eqnarray*}
Since the SI system of units accommodates quantities with the dimensions $q$  of electric charge, it is convenient at this point to express  (\ref{SecOrderMax}) as
\begin{eqnarray*}
	d\hindexMaxF{2}\,=\, 0, \qquad d \star \hindexMaxF{2}     = \widehat{\mathcal{J} }
\end{eqnarray*}
in terms of the dimensioned forms:
\begin{eqnarray*}
	\hindexMaxF{2}\,=\, \frac{ q_0} {L_0 \epsilon_0 }\,\indexMaxF{2} ,  \qquad          \widehat{\mathcal{J} }= \frac{ q_0} {L_0 \epsilon_0 }\, \mathcal{J}  
\end{eqnarray*}
where the real constant $q_0$  carries the dimensions of charge  and $\epsilon_0$ is the permittivity of free space with SI dimensions  $\text{q}^2\,\text{T}^2/ \text{M}\,\text{L}^3$. Then in units with charge q  measured in Coulombs,  Newtonian force (N) measured in Newtons  and length (L) in metres, $\epsilon_0=  8.8542 \times 10^{-12} \, \text{q}^2 \, \text{N}^{-1}\, \text{L}^{-2}$. At any spacetime point, in any local frame of reference defined by a unit time-like vector field  $V$ on $\M$, the charge density  $\rho^{ (V)} $ in Coulombs per cubic metre is then given by $  -  i_V\, \star{ \widehat{\mathcal{J}    }}$ and the associated electric current  $2$-form $J^{(V) }  $ in amps per square metre is given by $ c\,i_V \,  { \widehat{\mathcal{J}    }}  $ at that point.\\

Since $\indexMaxF{0}$ and $\indexMaxF{1}$ are source free solutions, assumed regular and singularity free, the net electric charge $Q[\Sigma_3]$  in any 3-dimensional   spacelike domain  $\Sigma_3 \subset S^3$   with boundary $\partial \Sigma_3 \neq 0 $,  is:
\begin{eqnarray*}
	Q[\Sigma_3] \,=\, \int_{ \Sigma_3 }\widehat{ \mathcal{J}} =    \int_{ \Sigma_3 } d\star \hindexMaxF{2}.
\end{eqnarray*}
This implies, by Stokes' theorem, that
\begin{eqnarray*}
	Q[\Sigma_3] \,=\,    \int_{\partial \Sigma_3 } \star\hindexMaxF{2}
\end{eqnarray*}
is the net electric flux emanating from the 2-surface $ \partial \Sigma_3 $ bounding  $\Sigma_3$. The value of $Q[\Sigma_3]$ may be positive or negative, implying the existence of localised charge fluctuations of  different polarities in different regions. However if one integrates $J$ over an entire {\it closed} space-like hypersurface, topologically  $S^3$, at any instant, then  $Q[S^3]=0$ since the 3-sphere has no boundary across which the electric flux can cross  (${\partial S^3 } =0$). Thus the {\it total electric charge} (induced by the scalar fields $ \indexScalarField{1}   $ and $ \indexMaxF{0}  $ at  order $\epsilon^2)$ must be zero. In any cosmological model the presence of locally charged regions in space seems to be an essential prerequisite for providing  scattering processes in a charged particle-field interpretation of the thermalisation mechanism between matter and radiation prior to any decoupling era where ionisation ceases.

\section{Thermal History within the New Paradigm}
The existence of a high temperature state of the early Universe has achieved almost universal recognition based on the accurate measurements of an extra-galactic omni-directional Planck distribution of electromagnetic radiation with a maximum intensity  in the microwave spectrum.  Assuming that such  a radiation distribution has evolved with the expansion of the Universe without change of shape since it escaped from the opaque plasma state, any cosmological model should accommodate the high precision COBE and Planck data \cite{PLANCK}. If 
\begin{eqnarray*}
	{\cal E}[\Sigma_{3}]=	V[\Sigma_{3}] \, \int_{0}^{\infty} \wh{U}_{\omega}(\tempT)\,d\omega 
\end{eqnarray*}
denotes the total thermal energy in any spatial domain $\Sigma_{3}$ with volume $V[\Sigma_{3}]$ in an early plasma state  then it has a  Planck spectrum with temperature $\tempT$ if
\begin{eqnarray*}
	\wh{U}_{\omega}(\tempT) \,=\, \frac{\hbar\omega^{3}}{\pi^{2}c^{3}}\;\frac{1}{\exp\!\(\frac{\hbar \omega}{k\tempT}\)-1}
\end{eqnarray*}
where $\hbar,k$ are the reduced Planck and Boltzmann constants respectively. Assuming that each harmonic frequency component in this distribution transports electromagnetic energy along a null geodesic of an isotropic background  LFRW metric $g$, one may exploit the ray-optics approximation for harmonic solutions to the source-free Maxwell equations \cite{plebanski_GRbook} to relate the frequencies or wavelengths of these components made by observers at different spacetime events $p$. In a cosmology with a LFRW metric a key feature is the existence of a preferred unit time-like  vector {\it field}  $V$ that can be used to define the  angular frequency $\omega_p=2 \pi\,\nu_{p} $ and wavelength $\lambda_p=2 \pi \,c/\omega_p$  of such solutions at any event $p$. Then
\begin{eqnarray*}
	\frac{\emitnu}{\obsnu} \,=\, \frac{g(V_{\text{e}},K_{\text{e}})}{g(V_{\text{o}},K_{\text{o}})} \,\equiv\, 1 + z_{\text{o,e}},
\end{eqnarray*}
in terms of the redshift parameter $z_{\text{o,e}}$. Here $e$ denotes an emission event where the tangent to the null geodesic is the null vector $K_e$ of a harmonic Maxwell wave with frequency $\nu_e$ in the cosmic frame $V_e$. A similar statement refers to the observation of the frequency $\nu_o$ at an observation point $o$ in the causal-future of $e$. With $V=\frac{1}{L_0}\frac{ \partial} { \partial t}$  it follows from (\ref{POLARMETRIC})  that
\begin{eqnarray}\label{ FREQRATIO }
	\frac{\nu_e}{\nu_o}=\frac{S(t_0)}{S(t_e)}
\end{eqnarray}
and 
\begin{eqnarray}\label{ epoch }
	z_{t_{0},t_{1}}=\frac{S(t_{0})}{S(t_{1})}-1=\sqrt{\left|\frac{\Sigma(t_{0})}{\Sigma(t_{1})}\right|} - 1
\end{eqnarray}
in terms of the general solution (\ref{SigmaSol}). Defining
\begin{eqnarray*}
	{U}_{\lambda}(\tempT) \,= \wh{U}_{\omega}(\tempT) \vert_{ \omega=\frac{2\pi c}{\lambda}  }
\end{eqnarray*}
and the change of variable
\begin{eqnarray}\label{ydef}
	y \,\equiv\, \frac{2\pi\hbar c}{\lambda k\tempT},
\end{eqnarray}
the distribution  ${U}_{\lambda}(\tempT) $ in wavelength becomes
\begin{eqnarray}\label{UNIVF}
	{U}_{y}(\tempT) \,d\lambda=\, \frac{k^{5}\tempT^{5}}{2\hbar^{4}\pi^{3}c^{4}}\,\mathcal{F}(y)\,d\lambda
\end{eqnarray}
in terms of  the dimensionless universal function
\begin{eqnarray}\label{UNIVF1}
	\mathcal{F}(y) \,=\, \frac{y^{5}}{e^{y}-1}.
\end{eqnarray}
Since the COBE and Planck data measure the {\it variation}  of ${U}_{y}(\tempT)$ with $\lambda_o$  with a definite temperature $\tempT_o\simeq 2.73^o\, K$ (after correcting for the detector's motion relative to the frame $V_o$) in agreement with (\ref{UNIVF}), it follows from (\ref{ydef}) that $y$ remains independent of epoch after decoupling, i.e.
 \begin{eqnarray*}
	\frac{ \lambda_o}{ \lambda_e }= \frac{\tempT_e   }{ \tempT_o  }
\end{eqnarray*}
and hence from (\ref{ FREQRATIO }):
 \begin{eqnarray*}
	S(t_e)=S(t_0)\,\frac{\tempT_o}{\tempT_e}.
\end{eqnarray*}
If the scale factor $S(t)$ is continuous monotonic-increasing with $t$ and $0<\tempT_{\text{o}}/\tempT_{\text{e}}<1$ then $t_{\text{e}}$ lies at a time between the initial singularity event and the current epoch $t_{0}$. An estimate of the temperature $\tempT_e$ when the Universe first became transparent to thermal radiation would give an estimate of the ratio $S(t_e)/S(t_0)$ since $\tempT_{o}$ is measured.
In the standard  cosmological model the value of $S(t_0)$ is estimated from the current densities  of matter and  radiation in a solution of the Friedman equations. In the cosmological  model discussed in this paper it is fixed by the solution (\ref{St})  using the current values of ${\cal H}(t_0)$ and $q(t_0)$ that fix the constants $ C_{s1}, C_{s2}$ and  $W$  assuming    $\Delta=1.257$. If one assumes a value $\tempT_e\simeq 4000^o K$  for the temperature at decoupling, based on a model for the physics of thermalisation in the hot plasma,  together with  a value $S(t_0)= 2.25\,\times 10^{27}$ defined by (\ref{S0})  one may estimate the value for the scale factor at decoupling to have been $  S(t_e)\simeq 1.54 \times 10^{24} $. Finally, denoting by $t_{BB}$ the dimensionless cosmic time of the geometric singularity, then  the dimensionless time interval between the big-bang and the decoupling era $ t_e -t_{BB} $,  is determined to be $ 7.78 \times 10^{19},$  i.e.  $8.23 \times 10^{3} $   years. 

\section{Extending the Fluctuation Model with Perturbative Scalar and Vector Minkowski Modes}
In the current epoch, with $t=t_{0}$,   many physical phenomena can be described in terms of field theories on a background  flat Minkowski spacetime. If one introduces new coordinates $\wh{T}$ and $\wh{R}$ (with dimensions of time and length respectively) by the substitutions $t=c\,\wh{T}/L_{0}$ and $r=\wh{R}/ S(t_0) L_0$ then in coordinate domains where $\wh{R}\ll  2 S(t_0) L_0$ the LFRW metric  (\ref{POLARMETRIC}) is well approximated by the Minkowski metric in standard spherical polar spatial coordinates:
\begin{eqnarray*}
	 g \,\approx\,  - c^2 \,d\wh{T} \otimes d \wh{T} + d\wh{R} \otimes d\wh{R} + \wh{R}^2 \, d\theta \otimes d\theta + \wh{R}^2\, \sin^{2}(\theta)\, d\phi \otimes d\phi.
\end{eqnarray*}
In this section the Hodge map $\star$ will, throughout, refer to the metric of this background, approximately flat, spacetime domain. In the previous sections we have shown that the solutions $\ScalarField=\ScalarField_0$ and $\ProcaA=\ProcaC\,e^1$ are exact solutions to  the equations (\ref{psi_eqn}) and (\ref{proca_eqn}) in a particular LFRW metric. However these  coupled non-linear {\it partial differential equations}  on $\M$ have  other solutions that may have relevance for the generation of local spatial inhomogeneities during the evolution of the Universe. If we assume that such solutions are perturbations of the exact cosmological solutions discussed above  then a linearisation of (\ref{psi_eqn}) and (\ref{proca_eqn}) will offer a perturbative approach for finding them. Thus we analyse the first $\epsilon$-order equations (\ref{FirstOrderPsi_1}), (\ref{FirstOrderProca})  for perturbative solutions about the zeroth $\epsilon$-order  solutions $ \indexScalarField{0} =\ScalarField_{0} ,  \indexProcaA{0} = \ProcaC\,e^{1} ,  \indexMaxF{0} =0$, i.e. the  uncoupled {\it linear partial differential equations}  for $ \indexScalarField{1}  $  and $  \indexProcaA{1}$ :
\begin{eqnarray}\label{GenKleinGordon}
	\star d \star d\indexScalarField{1} + \frac{W_1}{L_{0}^2}\,\indexScalarField{1} &=& 0
\end{eqnarray}
\begin{eqnarray}
	\begin{split}\label{GenProcaEqn}
		\star d \star d\indexProcaA{1} + \frac{W}{L_{0}^{2}}\,\indexProcaA{1} &\,=\; 0 \\[0.1cm]	
	\end{split}
\end{eqnarray}
where
\begin{eqnarray*}
	W_1 \,=\, \frac{1}{2}\zeta_{0}\,\Xi_{1}\ProcaC^{2}, \qquad	W \,=\, \zeta_{0}\,\Xi_{0}.
\end{eqnarray*}
Since the divergence operator $\delta\equiv \star d \star$ is nilpotent ($\delta^2=0$),  all solutions of  (\ref{GenProcaEqn}) satisfy 
\begin{eqnarray*}
	\delta \indexProcaA{1} \,=\, 0.
\end{eqnarray*}
It is instructive to compare  equations   (\ref{GenKleinGordon}), (\ref{GenProcaEqn}) with the classical Klein-Gordon and Proca field equations  in Minkowski spacetime. These were introduced historically to accommodate  short-range Yukawa-type static field (singular) solutions associated with  {\it  particles}  having a real positive rest-mass $m_0$ and $m_P$ respectively. Expressed in terms of their respective Compton wavelengths such equations take the form:
\begin{eqnarray}\label{KleinGordon}
	\star d \star d\indexScalarField{1} + \frac{m_{0}^{2}c^{2}}{\hbar^{2}}\,\indexScalarField{1} &=& 0
\end{eqnarray}
and
\begin{eqnarray}\label{ProcaEqn}
	\begin{split}
		\star d \star d\indexProcaA{1} + \frac{m_{\mbox{\tiny $P$}}^{2}c^{2}}{\hbar^{2}}\,\indexProcaA{1} &\,=\, 0 \\[0.1cm]
		\delta \indexProcaA{1} &\,=\, 0.	
	\end{split}
\end{eqnarray}
where
\begin{eqnarray*}
	m_{0}^{2} \,=\, \frac{W_{1}\hbar^{2}}{L_{0}^{2}c^{2}}\qquad\text{and}\qquad m_{\mbox{\tiny $P$}}^{2} \,=\, \frac{W\hbar^{2}}{L_{0}^{2}c^{2}}.
\end{eqnarray*}
One sees that when the real parameters $W$ and $W_1$ are strictly positive, (\ref{GenKleinGordon}) and (\ref{GenProcaEqn}) yield real values for $m_{0}$ and $m_{\mbox{\tiny $P$}}$  but admit  different types of  solutions  otherwise. Thus perturbative  corrections  to  the cosmological solution    $  \indexProcaA{0}    $ in the previous section require $W= - {\cal H}(0)^2\,\Delta^2 < 0 $ with $\Delta \in \real$ so do not admit  solutions with a real Proca rest-mass $m_P$. However  the sign of $\zeta_{0}$ is a-priori unconstrained and if $W_{1}$ is positive  then perturbative  corrections  to  the constant solution    $   \indexScalarField{0}   $ can describe a  perturbed scalar field associated with  particles of mass $m_0$ that could, in the absence of further constraints,  provide a dark-matter candidate since it has no direct coupling to the Maxwell field.\\

However the presence or absence of  static solutions to either  (\ref{GenKleinGordon}) or (\ref{GenProcaEqn}) does not imply the absence of  dispersive  propagating wave-packet  (radiative) solutions constructed by superposition of separable mode solutions to these wave equations.  \\
 
To illustrate this we   consider an interesting class of exact solutions in local spatial cylindrical polar coordinates $\{\wh{T}, \wh{r}, \wh{\phi}, \wh{Z}\}$ in which the Minkowski metric tensor takes the form: 
\begin{eqnarray*}
	\wh{g} \,=\,  - c^2 \,d\wh{T} \otimes d \wh{T} + d\wh{r} \otimes d\wh{r} + \wh{r}^2 \, d\wh{\phi} \otimes d\wh{\phi} + d\wh{Z}\,  \otimes d\wh{Z}.
\end{eqnarray*}
This choice of cylindrical polar coordinates will facilitate our discussion of particular exact solutions below.
In these coordinates, a  general integral representation of  {\it complex solutions}  for (\ref{GenKleinGordon}) may be constructed from a   Fourier superposition of time harmonic and spatial cylindrical Helmholtz modes:
\begin{eqnarray*}
	 \indexScalarField{w_{1}}(\wh{T},\wh{r},\wh{\phi},\wh{Z}) =\sum_{\text{modes}} \sum_{m=-\infty}^{\infty}\int_{s=0}^{\infty}\int_{\omega=-\infty}^{\infty} \wh{\ScalarField}(\omega,s,m)\,J_{m}(\wh{r}\,s)\,e^{-i\omega\wh{T} + ik(w_{1},\omega,s)\,\wh{Z} + i m\wh{\phi} }\,d\omega\,ds\quad
\end{eqnarray*}
in terms of the complex scalar Fourier amplitudes $\wh{\ScalarField}$, the parameter $w_{1}=W_{1}/L_{0}^2 \in \real$  and the complex branched dispersion relation:
\begin{eqnarray}\label{DREL}
	k(w_{1},\omega,s) \,=\, \sqrt{\, \frac{\omega^{2}}{c^{2}} - s^{2} - w_{1}\;}.
\end{eqnarray}
Similarly a Fourier representation of solutions for (\ref{GenProcaEqn}) takes the form
\begin{eqnarray*}
	 \indexProcaA{w}(\wh{T},\wh{r},\wh{\phi},\wh{Z}) =\sum_{\text{modes}} \sum_{m=-\infty}^{\infty}\int_{s=0}^{\infty}\int_{\omega=-\infty}^{\infty} \wh{\ProcaA}(\omega,s,m)\,J_{m}(\wh{r} \,s)\,e^{-i\omega \wh{T} + ik(w,\omega,s)\wh{Z} + im\wh{\phi} }\,d\omega\,ds\quad
\end{eqnarray*}
where
\begin{eqnarray*}
	\wh{\ProcaA}(\omega,s,m) &=& \wh{\alpha}(\omega,s,m)\(e^{3} - \frac{c\,k(w,\omega,s)}{\omega}\,e^{0}\)
\end{eqnarray*}
in terms of $w=W/L_{0}^2 \in \real$, the complex scalar Fourier amplitudes $\wh{\alpha}$, the 1-forms  $e^{0}=c\, d\wh{T}$, $e^{3}=d\wh{Z}$ and  the complex branched dispersion relation:
\begin{eqnarray}\label{DREL1}
	k(w,\omega,s) \,=\, \sqrt{\, \frac{\omega^{2}}{c^{2}} - s^{2} - w\;}
\end{eqnarray}
The distinct mode types in the mode summations for $\indexScalarField{w_{1}}$ and $ \indexProcaA{w}$ above arise  from the distinct contributions to the integrations over $\omega$ and $s$ for each integer $m$,  determined by the sign of the argument of the square root function in the dispersion relations (\ref{DREL}) and (\ref{DREL1}) respectively.\\

For each mode labelled by $m$  in the summations for $\indexScalarField{w_{1}}$ and $ \indexProcaA{w}$ above, the double integral over the half-plane $\omega\in [ -\infty , \infty] $, $ s\in [0, \infty]$ can be partitioned into distinct contributions  by branches of the hyperbolae $ k(w_1,\omega,s)=0$ and $ k(w,\omega,s)=0$ respectively. Contributions with $k$ real yield progressive modes  while those with $k$ pure imaginary   behave exponentially with respect to $\wh{Z}$. The contributing physical modes are those that are spatially ``evanescent'' for all  $\wh{Z}$.%
\BESSELFOURIERMODEDOMAINS{0.88}{%
	Light grey domains ($\omega^{2}/c^{2} -s^{2}-w > 0$) distinguish contributions to the Fourier integral representations yielding progressive modes from dark grey domains ($\omega^{2}/c^{2}-s^{2}-w < 0$) that contribute to evanescent modes.
	}
Figure~\ref{fig:besselfouriermodedomains} indicates how this partitioning depends on the sign of  $w_{1}$ for the dispersion relation (\ref{DREL}). A similar partition is determined by the branch points of the dispersion relation (\ref{DREL1}) where $w$ has replaced  $w_{1}$. In general the global  behaviour of solutions generated from such mode summations depends critically on the structure of the scalar Fourier amplitudes $\wh{\alpha}(\omega,s,m)$  and $\wh{\ScalarField}(\omega,s,m)$.\\

\subsection{Particular Local Fluctuations in Minkowski Spacetime}
The existence of particular examples of  {\it axially symmetric}    $(m=0)$ dispersive  complex wave-packet solutions of the equations (\ref{GenKleinGordon})  and (\ref{GenProcaEqn}) can be  directly demonstrated using the methods of pre-potentials \cite{brittingham,synge1956relativity,ziolkowski_JMP} and recently exploited to study the dynamic evolution of electromagnetic single-cycle laser pulses \cite{goto_lasers_JPA}. In terms of  {\it strictly  positive} arbitrary real constants $Q_{1},Q_{2}$ with dimensions of length, introduce the  dimensionless expression
\begin{eqnarray}\label{LAMBDA}
	\Lambda(\wh{T},\wh{r},\wh{Z}) &\equiv& \frac{\sqrt{ \wh{r}^{2} + [Q_{1}+i(\wh{Z}-c\wh{T})]\,[Q_{2}-i(\wh{Z}+c\wh{T})] \;}}{L_{0}}.
\end{eqnarray}	
Then  one may verify directly that a particular complex axially-symmetric non-singular dispersive (radiative)  pulse solution to (\ref{GenKleinGordon}) is given by the scalar fields:
\begin{eqnarray}
	\label{GenKGPulseSol1}	\indexScalarField{1}_{W_{1}}(\wh{T},\wh{r},\wh{Z}) &=& \frac{K_{1}\(   \sqrt{ W_{1} }    \,\Lambda(\wh{T},\wh{r},\wh{Z})\,\)}{\Lambda(\wh{T},\wh{r},\wh{Z})} \qquad \text{when  }  W_{1}>0 \\[0.2cm]
	\label{GenKGPulseSol2}	\indexScalarField{1}_{W_{1}}(\wh{T},\wh{r},\wh{Z}) &=& \frac{  {\cal C }\(   \sqrt{ -W_{1} }    \,\Lambda(\wh{T},\wh{r},\wh{Z})\,\)}{\Lambda(\wh{T},\wh{r},\wh{Z})} \qquad \text{when  }  W_{1} < 0
\end{eqnarray}
where $K_{1}$  is a modified second kind Bessel function    and    ${\cal C} $ is  either the Hankel function   $H^{(1)}_{1}$ or $H^{(2)}_{1}$. \\

It has been shown in \cite{ziolkowski_PRA,goto_lasers_NIMB} how solutions to certain linear scalar wave equations can be used to generate solutions to certain linear vector wave equations. A similar process is available here to generate a particular solution to (\ref{GenProcaEqn}) from the solutions (\ref{GenKGPulseSol1}) or (\ref{GenKGPulseSol2}) with $W_{1}$ replaced by $W$.  Thus for {\it any real}  $W\neq 0$,  complex   dispersive  $\nu$-chiral  vector wave-packet solutions to (\ref{GenProcaEqn}) are  given in terms of (\ref{GenKGPulseSol1}) or (\ref{GenKGPulseSol2}) by
\begin{eqnarray}\label{GenProcaPulseSol}
	\indexProcaA{1}_{\nu}(\wh{T},\wh{r},\wh{Z}) &=& \star d\(\frac{}{} \indexScalarField{1}_{W}(\wh{T},\wh{r},\wh{Z})\,\,\Pi_{\nu} \),
	\qquad \nu=\pm 1,0
\end{eqnarray}
where the closed ($ d\Pi_{\nu}=0 $)  and co-closed  ($\delta\Pi_{\nu}=0   $) complex chiral 2-forms $\Pi_{\nu}$ are given by
\begin{eqnarray*}
	\Pi_{\pm 1} \,=\, d(\,\wh{r} e^{\pm i\wh{\phi}}\,) \w c\, d\, \wh{T}, \qquad	\Pi_{0} \,=\, d \wh{Z} \w  c\, d \wh{T}.
\end{eqnarray*}
The detailed ``dispersive'' behaviour of these axially-symmetric complex wave-packets, as functions of  $(\wh{T},\wh{r},\wh{Z})$, is determined by the  ratio  $Q_1/Q_2$ and the parameters $W, W_1$. The real (or imaginary) parts  describe real  non-singular solutions to  (\ref{GenKleinGordon}) and  (\ref{GenProcaEqn}). By construction the expression (\ref{LAMBDA}) generates  solutions  with turning points at  particular values  when $\wh{T}=0$.  The location of these  turning points can easily be changed by shifting the origin of the spatial coordinates. Similarly the direction of the  $\wh{Z}$-axis of symmetry can be arbitrarily rotated to another direction \cite{visser_wavelets}. Those solutions involving the $K_{1}$ and $ H^{(2)}_{1}    $  Bessel functions exhibit bounded axially-symmetric wave-packet  profiles that spread in $\wh{r}$ and $\wh{Z}$ as $\wh{T}$ increases. By contrast those involving $H^{(1)}_{1}$ Bessel functions exhibit growing amplitudes as $\wh{T}$ increases and are unphysical solutions.\\

\PROCAMASSPLOTS{1}{$\{W,W_{1},\zeta_{0},\Xi_{0}\}$ correlations for solutions for the Proca and Klein-Gordon wave equations. }
Given the properties of the particular scalar and vector perturbations discussed in this section, one can summarise the following  criteria on the basic parameters for the perturbed cosmological model. We demand $W\equiv \zeta_{0}\,\Xi_{0} <0$ to be consistent with $q(0)<0$ and $\Xi_{1}\equiv   \Upot''(\indexScalarField {0})>0 $ for a potential minimum at $\indexScalarField{0}$. However solutions that admit Klein-Gordon and Proca  positive rest-mass solutions require $W_{1} > 0$ and $W>0$ respectively where $ W_{1}\equiv \zeta_{0}\,\Xi_{1} \,\Gamma^2/2$ with $\Gamma$ real.  Furthermore solutions with arbitrary non-zero values of  $W_{1}  $ and $W$  admit scalar field  and vector field physical wave-packet profiles  respectively. Since $ \zeta_{0} $ and $ \Xi_{0}\equiv\Upot(\indexScalarField {0})$ remain unconstrained we exhibit these conditions as designated  domains in the $ \Xi_{0} - \zeta_{0} $ plane in Figure~\ref{fig:procamassplots}.
Shaded domains in these figures:
\begin{enumerate}[label=(\roman*),leftmargin=2cm]
	\item   admit  construction of a viable cosmological model  from  $S(t) $ with one real root.
	\item	admit scalar perturbations $\indexScalarField{1}$ that satisfies the Klein-Gordon equation  (\ref{KleinGordon}) with a real rest-mass $m_0$.
	\item	admit  vector perturbations $\indexProcaA{1}$  satisfying the Proca equation (\ref{ProcaEqn}) with a real rest-mass $m_P$.
	\item	admit both (ii) and (iii).
	\item	admit vector perturbations $\indexProcaA{1}$  satisfying  a Proca equation with a real rest-mass  $m_P$ and spatially bounded dispersive wave packets (\ref{GenKGPulseSol1}).
	\item	admit  vector perturbations $\indexProcaA{1}$  described by spatially bounded dispersive wave packet solutions (\ref{GenProcaPulseSol}) and scalar perturbations $\indexScalarField{1}$  described by  solutions  with a real rest-mass $m_0$ and  spatially bounded dispersive wave packet solutions (\ref{GenKGPulseSol1}).
	\item   admit  vector perturbations $\indexProcaA{1}$  described by  solutions  with spatially bounded dispersive wave packets (\ref{GenProcaPulseSol})
			and  scalar perturbations $\indexScalarField{1}$  described by  solutions  with spatially bounded dispersive wave packets  (\ref{GenKGPulseSol1}). 
\end{enumerate}

These designated domains indicate that there exist real values of the couplings $\zeta_{0}$ and $\Xi_{0}$   permitting the construction of the  cosmological solution with $W<0$ discussed in this paper.  Such couplings then permit the construction of perturbed solutions $ \indexScalarField{1} $ and $\indexProcaA{1}  $ to the linear wave equations  (\ref{GenKleinGordon}) and  (\ref{GenProcaEqn}) in a background  domain that is   approximately Minkowskian. The former has particular solutions  characteristics of a Klein-Gordon solution with a real particle mass $m_{0}$  or an axially symmetric dispersive propagating wave packet while the latter has only axially symmetric dispersive propagating polarised vector wave packet solutions. Assuming that the history of such dispersive fluctuations  are currently detectable, an observational signature may reside in the  influence of a primordial vector field polarisation on visible matter (or electromagnetic fields)  or effects due to  a primordial massive Klein-Gordon field.\\

\section{Conclusions}{  }

In this article we have discussed a new paradigm for exploring a number of  puzzling aspects in modern cosmology and their implications for astrophysics and observable astronomy. In this paradigm we argue that an evolutionary description of the Universe  is best formulated in terms of a series of successive approximations based on a viable {\it cosmography} derived from  current observations with a minimum number of phenomenological constraints on the dynamics of the unobservable early Universe.  Consequently assumptions about the states of matter during such epochs in our paradigm are replaced  by a series of retro-dictions  from a coupled system of field equations with initial conditions  based on current data. We depart from many standard  cosmological models, with their use of different  isotropic fluid models in different epochs, by using a single anisotropic material fluid model in an  Einstein-Maxwell-vector-scalar-fluid system of master equations.   The Maxwell, vector and scalar  fields are coupled to gravity and themselves in such a way that an exact analytic approach to a cosmological solution for the spacetime metric, vector, scalar and Maxwell fields can be found without  the need to impose any a-priori equation of state for the material fluid. Instead, its equation of state in the cosmological sector is induced from the Einstein equation containing  a stress-energy-momentum tensor without a cosmological constant.  By establishing all solutions on a spacetime with the topology ${\cal I} \times S^3,  {\cal I}   \in  \real$ the spacetime metric falls within the LFRW class of metrics possessing spatial sections with topology $S^3$ and the  associated  scale factor  describes five distinct spacetime geometries. Motivated by the recent estimates of the  negative de-acceleration parameter we use the current value of the Hubble parameter to select a viable cosmology with a single singular state and an exponentially expanding scale factor. Its predicted history leads to the value  $13.2 \times 10^9 $ years for the age of the Universe and a predicted value of 1.48 for the (unmeasured) ``jerk'' parameter. Based on this history we have verified that, over an interval from the big-bang, including the current epoch, the Dominant-Energy, Weak-Energy and Null-Energy conditions are satisfied and only the Strong-Energy conditions are (weakly) violated.\\

By neglecting any back-reaction of the fields on the LFRW spacetime we have derived a series of coupled linear PDE's  that determine the vector, scalar and Maxwell field fluctuations  by the method of successive approximations. In this scheme one finds that,  to lowest {\it and} first order,  the primordial vector field remains ``dark'' (i.e. has no direct coupling to the Maxwell field at those orders). Furthermore we show that at second order the Maxwell field acquires an electric current source induced from lower order scalar fields and source-free Maxwell perturbations. We argue that these currents offer a potential mechanism for  initiating a thermalisation process between matter and electromagnetic radiation. Based on the COBE observation of  a microwave dominant Planck spectrum of cosmological origin  and a choice of $4000^{\circ}K$   for the temperature at radiation decoupling from matter,  our cosmological model history predicts  a time interval of  $8.23\times 10^3$ years between the big-bang and the decoupling era. \\

By restricting to  spacetime  domains where the local effects of gravity may be neglected we use the parameters  fixed  by cosmography and the coupling constants that enter into the master field equations to explore the relation of the vector and scalar perturbations to the solutions of the historic Proca and Klein-Gordon equations in Minkowski spacetime. We show  that there exist couplings  where both vector and scalar dispersive wave-packet solutions  arise  and where the scalar perturbation may give rise to Yukawa solutions associated with a massive  Klein-Gordon particle with rest mass given by
\begin{eqnarray*}
	m_{0}^{2} \,=\, \frac{\zeta_{0}\Xi_{1}\ProcaC^{2}\hbar^{2}}{2L_{0}^{2}c^{2}} \qquad (\zeta_{0}>0).
\end{eqnarray*}
In S.I. units ($L_{0}=1$), this has the value
\begin{eqnarray*}
	m_{0}^{2} \,=\, 0.616\times 10^{-84}\zeta_{0}\Xi_{1}\ProcaC^{2}.
\end{eqnarray*}
Dark matter searches are then circumscribed by the  values of the dimensionless parameters $\zeta_{0}, \Xi_0$ and $\Gamma$ which must be determined independently from other astrophysical observations. These include applications of the perturbation scheme in section 8 to gravitational lensing by galaxies, bounds on primordial magnetic fields,  CMBR anisotropies and dark-energy constraints.\\ 

Aside from such local perturbative features the global aspects of our model give rise to a number of novel electromagnetic solutions that owe their existence to the three-sphere topology of space. The symmetry of the electric and magnetic bi-pole solutions discussed in section 4 may have relevance to   charge conjugation  and parity inversion symmetry in the early Universe while primordial
extragalactic magnetic fields may owe their existence to evolving ``Hopfion-like'' solutions with magnetic helicity.\\

Further investigations based on the Einstein-vector-scalar-Maxwell-fluid paradigm discussed above may have implications for other more challenging problems in cosmology and astrophysics and will be discussed elsewhere. However, it is important to recognise that to confront our model with such astrophysical phenomena, the latest experimental data \cite{PLANCK} needs to be analysed beyond the standard paradigm.


\section*{Acknowledgements}
RWT is grateful to the University of Bolton for hospitality and to STFC (ST/G008248/1) and EPSRC (EP/J018171/1) for support. MA and JLT are supported by the research grant from the Spanish Ministry of Economy and Competitiveness ESP2017-86263-C4-3-R. All authors are grateful to Ed Copeland and Clive Speake for useful discussions.

\appendix
\setcounter{section}{1}
\section*{Appendix}
In this Appendix  the stereographic mapping employed in section 3 in the construction of a  spacetime LFRW metric tensor, in terms of Maurer-Cartan forms on $S^3$, is described.\\

An  $SO(3)$ group element $U$ can be expressed in terms of the Pauli matrices $\{\sigma_1, \sigma_2, \sigma_3 \}$ 
and group coordinates   $\{\alpha_1,  \alpha_2, \alpha_3 \}$     as
\begin{eqnarray*}
	U \,=\, \exp\[i(\alpha_{1}\sigma^{1} + \alpha_{2}\sigma^{2} + \alpha_{3}\sigma^{3})\]	.
\end{eqnarray*}
The three pure imaginary Maurer-Cartan 1-forms
\begin{eqnarray*}
	\Thetat^{k} \,=\, \frac{1}{2}\text{tr}(\sigma^{k}\,U^{-1}dU) \qquad (k=1,2,3)
\end{eqnarray*}
in the real coordinate cobasis $\{d\alpha_1,  d\alpha_2,d\alpha_2 \}$   are  then:
\begin{eqnarray*}
	\Thetat^{k} &=& \frac{i}{\alpha^{3}}\sum_{j=1}^{3}\Thetat^{k}_{j}\,d\alpha_{j}
\end{eqnarray*}
where 
\begin{eqnarray*}
	\Thetat^{1}_{1} &=& (\alpha_{2}^{2} + \alpha_{3}^{2})\cos(\alpha)\sin(\alpha) + \alpha_{1}^{2}\,\alpha \\[0.1cm]
	\Thetat^{1}_{2} &=& \alpha_{1}\,\alpha_{2}\(\,\alpha - \cos(\alpha)\sin(\alpha)\,\) - \alpha_{3}\,\alpha\sin^{2}(\alpha) \\[0.1cm]
	\Thetat^{1}_{3} &=& \alpha_{1}\,\alpha_{3}\(\,\alpha - \cos(\alpha)\sin(\alpha)\,\) - \alpha_{2}\,\alpha\sin^{2}(\alpha)
\end{eqnarray*}
\begin{eqnarray*}
	\Thetat^{2}_{1} &=& \alpha_{1}\,\alpha_{2}\(\,\alpha - \cos(\alpha)\sin(\alpha)\,\) + \alpha_{3}\,\alpha\sin^{2}(\alpha)\\[0.1cm]
	\Thetat^{2}_{2} &=& (\alpha_{1}^{2} + \alpha_{3}^{2})\cos(\alpha)\sin(\alpha) + \alpha_{2}^{2}\,\alpha \\[0.1cm]
	\Thetat^{2}_{3} &=& \alpha_{2}\,\alpha_{3}\(\,\alpha - \cos(\alpha)\sin(\alpha)\,\) - \alpha_{1}\,\alpha\sin^{2}(\alpha)
\end{eqnarray*}
\begin{eqnarray*}
	\Thetat^{3}_{1} &=& \alpha_{1}\,\alpha_{3}\(\,\alpha - \cos(\alpha)\sin(\alpha)\,\) - \alpha_{2}\,\alpha\sin^{2}(\alpha)\\[0.1cm]
	\Thetat^{3}_{2} &=& \alpha_{2}\,\alpha_{3}\(\,\alpha - \cos(\alpha)\sin(\alpha)\,\) + \alpha_{1}\,\alpha\sin^{2}(\alpha) \\[0.1cm]
	\Thetat^{3}_{3} &=& (\alpha_{1}^{2} + \alpha_{2}^{2})\cos(\alpha)\sin(\alpha) + \alpha_{3}^{2}\,\alpha
\end{eqnarray*}
and 
\begin{eqnarray*}
	\alpha^{2} \,\equiv\, \alpha_{1}^{2} + \alpha_{2}^{2} + \alpha_{3}^{2}. 
\end{eqnarray*}

In these equations the coordinates are defined in the intervals $ -\pi < \alpha_{j} < \pi  $  and $ 0 \le \alpha <\pi  $. To facilitate the construction of the LFRW metric from the Maurer-Cartan forms  $\Thetat^{k}$  on $S^3$ 
we introduce the stereographic coordinate transformation  $ \{\alpha_1,  \alpha_2, \alpha_2 \} \mapsto  \{\xi_1,  \xi_2, \xi_3 \}  $  :
\begin{eqnarray}\label{StereoCoordT}
	\alpha_{1} \,=\, \(\frac{\xi^{2}-1}{\xi^{2}+1}\)\chi, \quad \alpha_{2} \,=\, \(\frac{2\xi_{3}}{\xi^{2}+1}\)\chi, \quad \alpha_{3} \,=\, \(\frac{2\xi_{2}}{\xi^{2}+1}\)\chi
\end{eqnarray}
where
\begin{eqnarray*}
	\chi &\equiv& \frac{\arctan\( \sqrt{1-\omega^{2}}, \omega \)}{\sqrt{1-\omega^{2}}},
\end{eqnarray*}
and
\begin{eqnarray*}
	\xi^2\equiv\xi_1^2 + \xi_2^2 + \xi_3^2 ,  \qquad  \omega\equiv \frac{ 2\,\xi_1} {\xi^2 +1}.
\end{eqnarray*}
The two-argument function $\arctan(y,x)$  computes the principal value of the argument of the complex number $x+iy$ such that $\arctan(y,x)\in(-\pi,\pi]$. In the $\{\xi_1,  \xi_2, \xi_3 \} $ chart the Maurer-Cartan 1-forms $\{\Thetat^{1},\Thetat^{2},\Thetat^{3}\}$ become: 
\begin{eqnarray*}
	 \Theta^{1} &=& \frac{4i}{(1+\xi_{1}^{2}+\xi_{2}^{2}+\xi_{3}^{2})^{2}}\(  \frac{1}{2}(1+\xi_{1}^{2}-\xi_{2}^{2}-\xi_{3}^{2})d\xi_{1} + (\xi_{1}\xi_{2}+\xi_{3})d\xi_{2} + (\xi_{1}\xi_{3}-\xi_{2})d\xi_{3} \) \\[0.4cm]
	 \Theta^{2} &=& \frac{4i}{(1+\xi_{1}^{2}+\xi_{2}^{2}+\xi_{3}^{2})^{2}}\(  (\xi_{1}\xi_{2}-\xi_{3})d\xi_{1} + \frac{1}{2}(1+\xi_{2}^{2}-\xi_{1}^{2}-\xi_{3}^{2})d\xi_{2} + (\xi_{2}\xi_{3}+\xi_{1})d\xi_{3} \) \\[0.4cm]
	 \Theta^{3} &=& -\frac{4i}{(1+\xi_{1}^{2}+\xi_{2}^{2}+\xi_{3}^{2})^{2}}\(  (\xi_{1}\xi_{3}+\xi_{2})d\xi_{1} + (\xi_{2}\xi_{3}-\xi_{1})d\xi_{2} + \frac{1}{2}(1+\xi_{3}^{2}-\xi_{1}^{2}-\xi_{2}^{2})d\xi_{2}\) .
\end{eqnarray*}
In these equations the  coordinates are defined in  the intervals $ -\infty < \xi_{j} < \infty  $  and cover $S^3$ minus the point where $\xi^2\rightarrow\infty$ or, under the transformation: 
\begin{eqnarray*}
	\left\{\xi_{1}\,=\, \frac{1}{2}r\sin(\theta)\cos(\phi),\quad \xi_{2}\,=\, \frac{1}{2}r\sin(\theta)\sin(\phi),\quad \xi_{3}\,=\, \frac{1}{2}r\cos(\theta)\right\},
\end{eqnarray*}
at the point where $r^{2}\rightarrow\infty$.

\bibliographystyle{unsrt}
\bibliography{VecScalarCosmology}

\begin{thebibliography}{10}

\bibitem{ijjas_pop}
A.~Ijjas, P.~J. Steinhardt, and A.~Loeb.
\newblock Pop goes the universe.
\newblock {\em Scientific American}, 316(2):32--39, 2017.

\bibitem{ijjas_inflation}
A.~Ijjas, P.~J. Steinhardt, and A.~Loeb.
\newblock Inflationary paradigm in trouble after {P}lanck2013.
\newblock {\em Physics Letters B}, 723(4--5):261--266, 2013.

\bibitem{milgrom}
M.~Milgrom.
\newblock A modification of the {N}ewtonian dynamics as a possible alternative
  to the hidden mass hypothesis.
\newblock {\em Astrophys. J.}, 270:365--370, 1983.

\bibitem{bekenstein}
J.~D. Bekenstein.
\newblock Relativistic gravitation theory for the modified newtonian dynamics
  paradigm.
\newblock {\em Phys. Rev. D}, 70(8):083509, 2004.

\bibitem{thorne_primordial}
K.~S. Thorne.
\newblock Primordial element formation, primordial magnetic fields, and the
  isotropy of the {U}niverse.
\newblock {\em The Astrophysical Journal}, 148:51, 1967.

\bibitem{visser_jerk}
M.~Visser.
\newblock Jerk, snap and the cosmological equation of state.
\newblock {\em Classical and Quantum Gravity}, 21(11):2603, 2004.

\bibitem{plebanski_GRbook}
J.~Plebanski and A.~Krasinski.
\newblock {\em An {I}ntroduction to {G}eneral {R}elativity and {C}osmology}.
\newblock Cambridge {U}niversity {P}ress, Cambridge., 2006.

\bibitem{weinberg_cosmology}
S.~Weinberg.
\newblock {\em Cosmology}.
\newblock Oxford {U}niversity {P}ress, Oxford, 2008.

\bibitem{guth_inflation}
A.~H. Guth.
\newblock Inflation and eternal inflation.
\newblock {\em Physics Reports}, 333:555--574, 2000.

\bibitem{eckart_varhydro}
C.~Eckart.
\newblock Variation principles of hydrodynamics.
\newblock {\em The Physics of Fluids}, 3(3):421--427, 1960.

\bibitem{seliger_varCM}
R.L. Seliger and G.~B. Whitham.
\newblock Variational principles in continuum mechanics.
\newblock {\em Proc. Royal Soc. A: Math. Phys. Sci.}, 305(1480):1--25, 1968.

\bibitem{karlovini_stars}
M.~Karlovini and L.~Samuelsson.
\newblock Elastic stars in general relativity: {I}. {F}oundations and
  equilibrium models.
\newblock {\em Classical and Quantum Gravity}, 20(16):3613, 2003.

\bibitem{velten_supernovae}
H.~Velten, S.~Gomes, and V.~C. Busti.
\newblock Gauging the cosmic acceleration with recent type ia supernovae data
  sets.
\newblock {\em Physical Review D}, 97(8):083516, 2018.

\bibitem{alsabti_handbook}
A.~W. Alsabti and P.~Murdin.
\newblock {\em Handbook of {S}upernovae}.
\newblock Springer, Cham, Switzerland, 2017.

\bibitem{bass_photonmass}
L.~Bass and E.~Schrodinger.
\newblock Must the photon mass be zero?
\newblock {\em Proc. Royal Soc. A: Math. Phys. Sci.}, 232(1188):1--6, 1955.

\bibitem{achour_sphharm}
J.~Ben Achour, E.~Huguet, J.~Queva, and J.~Renaud.
\newblock Explicit vector spherical harmonics on the 3-sphere.
\newblock {\em Journal of Mathematical Physics}, 57(2):023504, 2016.

\bibitem{lindblom_harmonics}
L.~Lindblom, N.~W. Taylor, and F.~Zhang.
\newblock Scalar, vector and tensor harmonics on the three-sphere.
\newblock {\em General Relativity and Gravitation}, 49(11):139, 2017.

\bibitem{alertz_EMRW}
B.~Alertz.
\newblock Electrodynamics in {R}obertson-{W}alker spacetimes.
\newblock {\em Ann. Inst. H. Poincar\'{e} Phys. Th\'{e}or.}, 53(3):319--342,
  1990.

\bibitem{kopinski_GRG}
J.~Kopi{\'n}ski and J.~Nat{\'a}rio.
\newblock On a remarkable electromagnetic field in the {E}instein {U}niverse.
\newblock {\em Gen. Rel. Grav.}, 49(6):81, 2017.

\bibitem{berger_maghel}
A.~M. Berger and G.~B. Field.
\newblock The topological properties of magnetic helicity.
\newblock {\em Journal of Fluid Mechanics}, 147:133--148, 1984.

\bibitem{irvine_linkedlight}
W.~T.~M. Irvine and D.~Bouwmeester.
\newblock Linked and knotted beams of light.
\newblock {\em Nature Physics}, 4(9):716, 2008.

\bibitem{ranada_maghel}
A.~F. Ranada.
\newblock On the magnetic helicity.
\newblock {\em European Journal of Physics}, 13(2):70, 1992.

\bibitem{hazeltine_anisoplasma}
R.~D. Hazeltine, S.~M. Mahajan, and P.~J. Morrison.
\newblock Local thermodynamics of a magnetized, anisotropic plasma.
\newblock {\em Physics of Plasmas}, 20(2):022506, 2013.

\bibitem{visser_cosmography}
M.~Visser.
\newblock Cosmography: {C}osmology without the {E}instein equations.
\newblock {\em General Relativity and Gravitation}, 37(9):1541--1548, 2005.

\bibitem{visser_energyconds}
M.~Visser and C.~Barcelo.
\newblock Energy conditions and their cosmological implications.
\newblock In {\em Cosmo-99}, pages 98--112. World {S}cientific, Singapore.,
  2000.

\bibitem{daly_model}
R.~A. Daly and S.~G. Djorgovski.
\newblock A model-independent determination of the expansion and acceleration
  rates of the {U}niverse as a function of redshift and constraints on dark
  energy.
\newblock {\em The Astrophysical Journal}, 597(1):9, 2003.

\bibitem{PLANCK}
N.~Aghanim et~al.
\newblock Planck 2018 results. {VI}. {C}osmological parameters.
\newblock {\em arXiv:1807.06209}, 2018.

\bibitem{brittingham}
J.~N. Brittingham.
\newblock Focus waves modes in homogeneous {M}axwell’s equations:
  {T}ransverse electric mode.
\newblock {\em J. Appl. Phys.}, 54(3):1179--1189, 1983.

\bibitem{synge1956relativity}
J.~L. Synge.
\newblock {\em Relativity: the {S}pecial {T}heory}.
\newblock North-Holland Publishing Company, Amsterdam, 1956.

\bibitem{ziolkowski_JMP}
R.~W. Ziolkowski.
\newblock Exact solutions of the wave equation with complex source locations.
\newblock {\em J. Math. Phys.}, 26(4):861--863, 1985.

\bibitem{goto_lasers_JPA}
S.~Goto, R.~W. Tucker, and T.~J. Walton.
\newblock The dynamics of compact laser pulses.
\newblock {\em J. Phys. A: Math. Theor.}, 49(26):265203, 2016.

\bibitem{ziolkowski_PRA}
R.~W. Ziolkowski.
\newblock Localized transmission of electromagnetic energy.
\newblock {\em Phys. Rev. A}, 39(4):2005, 1989.

\bibitem{goto_lasers_NIMB}
S.~Goto, R.~W. Tucker, and T.~J. Walton.
\newblock Classical dynamics of free electromagnetic laser pulses.
\newblock {\em Nucl. Instr. Meth. Phys. Res. B}, 369:40--44, 2016.

\bibitem{visser_wavelets}
M.~Visser.
\newblock Physical wavelets: {L}orentz covariant, singularity-free, finite
  energy, zero action, localized solutions to the wave equation.
\newblock {\em Physics Letters A}, 315(3--4):219--224, 2003.

\end{thebibliography}

\end{document}